\documentclass{aa}

\usepackage{graphicx,amsmath,txfonts}
\usepackage{caption,subcaption,bm}
\usepackage[dvipsnames]{xcolor}
\usepackage{tablefootnote}
\usepackage{esint}
\usepackage{nicefrac}
\usepackage{subcaption}

\newcommand{\etaO}{\eta_{\rm O}}
\newcommand{\etaH}{\eta_{\rm H}}
\newcommand{\etaA}{\eta_{\rm A}}

\newcommand{\Mdota}{\dot{M}_{\rm a}}

\newcommand\Q{\mathcal{Q}^{\nicefrac{+}{-}}}
\newcommand\Qirr{\mathcal{Q}^+_{\rm irr}}
\newcommand\Qpdr{\mathcal{Q}^{\nicefrac{+}{-}}_{\rm pdr}}

\newcommand\eR{e_{\rm r}}

\newcommand\kP{\kappa_{\rm P}}

\newcommand\aR{a_{\rm R}}

\newcommand\yr{\,\rm yr}

\newcommand\au{\,\rm au}

\newcommand\Msun{\,M_\odot}
\newcommand\Msunyr{\,M_\odot\,{\rm yr}^{-1}}

\newcommand\betap{\beta_{\rm P}}

\newcommand{\ee}[1]{\times 10^{#1}}

\usepackage{color,soul,ulem}

\definecolor{dark-red}{rgb}{0.75, 0.00, 0.00}
\definecolor{hlcolor}{rgb}{1.00, 0.94, 0.92}\sethlcolor{hlcolor}
\definecolor{purple}{RGB}{148,0,211}

\renewcommand\emph[1]{\textit{#1}}

\usepackage{xspace}
\newcommand{\nirv}{\textsc{nirvana-iii}\xspace}

\newcommand{\V}{\mathbf{v}}
\newcommand{\B}{\mathbf{B}}
\newcommand{\hatb}{\mathbf{\hat{b}}}
\newcommand{\J}{\mathbf{J}}
\newcommand{\E}{\mathbf{\mathcal{E}}}

\newcommand{\vK}{v_{\rm K}}
\newcommand{\vB}{v_{\rm B}}

\newcommand{\vR}{v_{\rm R}}
\newcommand{\vph}{v_{\rm \phi}}
\newcommand{\vz}{v_{\rm z}}

\newcommand{\BR}{B_{\rm R}}
\newcommand{\Bph}{B_\phi}
\newcommand{\Bz}{B_{\rm z}}

\newcommand{\R}{\mathcal{R}}
\newcommand{\M}{\mathcal{M}}

\newcommand{\haa}{\textit{h-aa}\xspace}
\newcommand{\hf}{\textit{h-f}\xspace}
\newcommand{\ha}{\textit{h-a}\xspace}

\newcommand{\rhogd}{\rho_{\rm gap}/\rho_{\rm disk}}

\begin{document}

   \title{Global Hall-magnetohydrodynamic simulations of transition disks}

   \author{
     Eleftheria~Sarafidou\inst{1,2}
     \and
     Oliver~Gressel\inst{1}
     \and
     Barbara~Ercolano\inst{3,4}
   }

   \institute{
     Leibniz-Institut f\"ur Astrophysik Potsdam (AIP),
     An der Sternwarte 16, 14482, Potsdam, Germany
     \and
     Institut f\"ur Physik und Astronomie, Universit\"at Potsdam,
     Karl-Liebknecht-Str. 24/25, 14476 Golm, Germany
     \and
     Universit\"ats-Sternwarte, Ludwig-Maximilians-Universit\"at M\"unchen,
     Scheinerstr. 1, 81679 M\"unchen, Germany
     \and
     Exzellenzcluster “Origins”, Boltzmannstr. 2, D-85748 Garching, Germany
     \\ \email{elsarafidou@aip.de}
   }

   \date{Received month xx, 2024; accepted month xx, 2024}


   \abstract
   {Transition disks (TDs) are a type of protoplanetary disk characterized by a central dust and gas cavity. The processes behind how these cavities are formed and maintained, along with their observed high accretion rates of $10^{-8}$--$10^{-7} \Msunyr$, continue to be subjects of active research.}
   {This work aims to investigate how the inclusion of the Hall effect (HE) alongside Ohmic resistivity (OR) and ambipolar diffusion (AD) affects the structure of the TD. Of key interest is the dynamical evolution of the cavity and whether it can indeed produce transonic accretion, as predicted by theoretical models in order to account for the observed high accretion rates despite the inner disk's low density.}
   {We present our results of 2D axisymmetric global radiation magnetohydrodynamic (MHD) simulations of TDs for which all three non ideal MHD effects are accounted. We used the \nirv fluid code and initialized our model with a disk cavity reaching up to $R=8\au$ with a density contrast of $10^5$. We performed three runs, one with only OR and AD, and one for each of the two configurations that arise when additionally including the HE, that is, with the field aligned (anti-aligned) with respect to the rotation axis.}
   {For all three runs, our models maintain an intact inner cavity and an outer standard disk. MHD winds are launched both from the cavity and from the disk. Notably, when the HE is included, ring-like structures develop within the cavity. We moreover obtain accretion rates of $3$--$8\ee{-8} \Msunyr$, comparable to typical values seen in full disks. Importantly, we clearly observe (tran)sonic accretion $(v_{\rm acc} \gtrsim c_{\rm s})$ in the cavity. Additionally, outward magnetic flux transport occurs in all three runs.}
   {}

   \keywords{
     accretion, accretion discs -- magnetohydrodynamics -- protoplanetary discs -- stars: pre-main-sequence
   }

   \maketitle
%


\section{Introduction} \label{sec:intro}


Conventionally, transition disks (TDs) are protoplanetary disks (PPDs) that have inner dust cavities, as indicated by a deficit in their spectral energy distribution \citep[SED,][]{espaillat2014}. Initially, the term transition was intended to illustrate a shift from Class II to Class III, but subsequent studies have proposed that the cavity could also represent a morphological aspect of the disk \citep[see, e.g., the discussions in][]{wang2017,ercolano2017,vanderMarel2023}.

The source of these cavities remains an open question -- and proposed mechanisms are likely not exclusive \citep{vanderMarel2023}. One of the primary and most frequently explored mechanisms is gap formation by an embedded companion or planet \citep{marsh1992,rosotti2016,price2018} -- but only recently has this indeed been observed \citep[e.g.,][]{kepler2018}.
Another mechanism that has extensively been considered is photoevaporation, in a combination with viscous evolution \citep{clarke2001,alexander2014,ercolano2017,picogna2019}. Additionally, dust migration driven by radiation pressure has been proposed as a contributing factor -- although no consensus has been reached \citep[e.g.,][]{dominik_dullemond2011,owen_kollmeier2019,bi_fung2022}.

Apart from the dust cavity, there are clear indications of the presence of gas within the dust-depleted region \citep{najita2007,manara2014}. While the amount of gas in the cavity is not yet certain, recent studies suggest a drop of 2--4 orders of magnitude in the gas surface density profile \citep{vanderMarel2018}. Despite the uncertainty in the origin and evolution of the cavities, one recently emerging characteristic is that the accretion rate in a large number of TDs appears to be comparable to that of full disks \citep{manara2014,owen2016}. Various explanations have been proposed for this phenomenon. One possibility is transonic gas accretion \citep{wang2017}, where gas crosses the cavity. It has also been proposed that the high accretion rates are a result of a hidden mass reservoir in the innermost region of the disc. Models that consider this are usually also incorporating a dead zone  \citep{garate2022}; however detailed modeling suggests that a pressure bump does not always occur at the inner side of a dead zone \citep{delage2022}.

The majority of the models mentioned above use the $\alpha$-prescription \citep{shakura1973} for a viscously evolving disc. While the question of disk evolution and whether it is governed by turbulent transport or MHD winds is still open \citep[see][for a recent review]{manara2023}, it is increasingly clear that magnetic fields play an important role in the evolution of the system: On the one hand, Ohmic resistivity (OR) and ambipolar diffusion (AD) have been shown to severely suppress the magnetorotational instability (MRI) \citep{bai2011,gressel2015}, while on the other hand, magnetohydrodynamic (MHD) winds provide an effective alternative for angular momentum transport and gas accretion \citep{bai2017global,bethune2017,gressel2020}.

Additionally there is the added complexity of the Hall effect (HE). The HE essentially reorganizes the magnetic field topology with vastly different outcomes depending on the magnetic field polarity, ranging from suppressing turbulence or instability \citep{wardle_salmeron2012} to introducing a new one \citep[Hall shear instability,][]{kunz2008}. It has also been shown that the HE can enhance the radial and toroidal magnetic fields, thereby increasing the accretion rate through the laminar Maxwell stress \citep{bai2017hall, sarafidou2024}.

Relatively few models of transition disks with magnetic fields exist. Early on, \citet{combet2008} presented a theoretical model of a magnetic wind-driven inner disk via what they termed a jet-emitting disk (JED), demonstrating that it can indeed sustain accretion rates similar to the outer turbulent disk -- a situation reminiscent of transitional disks, as they commented upon. Later on, secular evolution models of (partially) wind-driven accretion allowed \citet{suzuki2016} to find that a surface-density-dependent torque can indeed create and sustain a central cavity. Based on static wind models in the framework of a weakly ionized plasma, \citet{wang2017} took into account the magnetic diffusivities in a TD and showed that they indeed allow for the launching of a magnetic wind -- and furthermore that this would lead to transonic gas accretion. Finally, \citet{martel_lesur2022} performed the first self-consistent numerical model of an accreting TD with a non ideal MHD disk wind. They confirmed that such a model cannot only maintain a cavity over thousands of orbits but also predict high accretion rates, indeed comparable with the sound speed within the cavity.

In this work, we present the first radiative non ideal MHD (RMHD) model of a TD, featuring several novel aspects: we include simplified stellar irradiation, diffuse re-radiation \citep{gressel2017}, as well as far ultra-violet (FUV) heating via a simple thermochemistry model \citep{gressel2020}. We moreover feature all three non ideal MHD effects, with diffusivities calculated depending on local properties of the gas and shielded exterior ionizing sources \citep{gressel2015,gressel2020}. Our goal is to examine which of the previously reported results hold under this more complex approach and to identify any new dynamical behaviors that may arise.

The remainder of this paper is organized as follows: In Section~\ref{sec:methods} we present the model equations, physical quantities and the numerical setup. Then, in Section~\ref{sec:results}, we proceed and present the results of our runs. We discuss the accretion rate (and its origin), the self organization of the flow in the cavity region, as well as the issue of magnetic flux transport. In Section~\ref{sec:discussion}, we discuss our results and compare them with existing work. Finally, we present our conclusions in Section~\ref{sec:conclusions}.

\section{Methods} \label{sec:methods}

We perform radiation MHD simulations of the inner regions of PPDs, refining the non ideal MHD disk model of \citet{gressel2020}, that featured stellar irradiation and a simplified thermochemistry. As the most notable improvement over that work, we now also include the Hall effect in the induction equation, which is implemented by means of the framework presented in \citet{krapp2018}. Additionally, since we aim to study TDs, we have introduced a cavity --that is, a drop of several orders of magnitude in the gas surface density-- in our disk. Note that unlike \cite{sarafidou2024} we do not include X-ray photoionization, in order to keep continuity with \cite{gressel2020}

Our simulations are 2D axisymmetric with a spherical-polar coordinate system $(r, \theta, \phi)$ - radius, colatitude, and azimuth, respectively. We are primarily interested in the inner regions of the PPD where the disk is considered to be laminar under typical conditions. Our computational domain covers an area of $r\in(0.75, 22.5)\au$, $\theta\in(0, \pi)$ with a standard grid resolution of $N_r\times N_{\theta} = 320\times 256$ cells. All simulations were performed with the single-fluid \nirv code, which is built around a standard second-order accurate finite volume scheme \citep{ziegler2004,ziegler2016}. We use a modified version of the \nirv code that has been described in detail in \citet{gressel2020}. To obtain the interface states entering the magnetohydrodynamic fluxes, we use the HLLD approximate Riemann solver by \citet{miyoshi2005}. The Hall term is added via operator-splitting, which moreover allows sub-stepping in time when required. The update due to the Hall electromotive force is done using the Hall Diffusion Scheme (HDS) of \citet{osullivan2006,osullivan2007}.


\subsection{Equations of motion}

We solve the equations of motion for mass-, momentum- and total energy-density in conservation form along with the induction equation for the magnetic field, that is,
\begin{eqnarray}
\partial_t\rho+\nabla\!\cdot(\rho\V) & = & 0\,,
\nonumber\\
\partial_t(\rho\V)+\nabla\!\cdot\big[\,\rho\V\V
+P \mathbf{I}-\B\B/\mu_0\,\big] & = & \!-\rho\nabla\Phi\,,
\nonumber\\
\partial_t e + \nabla\!\cdot
\big[\, (e+P)\V - (\V\cdot\B)\B/\mu_0\, \big]
& = & \!-\rho(\nabla\Phi)\!\cdot\V + \nabla\!\cdot\!\mathcal{S} + \Q \,,
\nonumber\\
\partial_t \B & = & - \nabla\times\E\,,
\nonumber
\end{eqnarray}
where the total pressure, $P$, is defined as the sum of the gas and the magnetic pressure, $B^2/2\mu_0$. Finally, $-\rho\nabla\Phi$ denotes the gravity force of the central star.

\subsubsection{Non ideal MHD} 

In the above energy equation, we have a divergence of the Poynting flux, $\mathcal{S} \equiv \E\times\B/\mu_0$, with $\E$ representing the electromotive force responsible for the evolution of the magnetic field. It takes into account induction, as well as the three non ideal MHD effects, that is,
\begin{equation}
        \E = -\V\times\B + \mu_0\left[\;\etaO\J + \etaH \J\times\hatb - \etaA(\J\times\hatb)\times\hatb\;\right]\,, \label{eq:emf}
\end{equation}
where  $\etaO, \etaA$ and $\etaH$ denote the diffusion coefficients of Ohmic resistivity, ambipolar diffusion and Hall effect, respectively, $\J = (1/\mu_0)\nabla\times\B$ is the electric current density, $\hatb$ the unit vector of the magnetic field, and $\mu_0$ the magnetic permeability.

For the dynamically evolving diffusion coefficients $\etaO$, $\etaH$, and $\etaA$, we use a lookup table based on \citet{ilgner_nelson2006} and \citet{desch2015}; details regarding the ionization model can be found in Section~2.5 of \citet{gressel2020}. Moreover, in order to avoid severe timestep constrains we limit $\etaA$ and $\etaH$ such that  $\Lambda_{\rm A}\,\betap \geq 10^{-3}$ and $\Lambda_{\rm H}\,\betap \geq 10^{-2}$, respectively.\footnote{$\Lambda$ denotes the associated Elsasser number, $\Lambda_{\rm A/H} \equiv B^2/(\rho\mu_0\,\Omega\,\eta_{\rm A/H})$, and $\beta_{\rm p}\equiv 2\mu_0\,p / B^2$ the plasma parameter. Combining the two expressions yields $2p/(\rho\Omega\,\eta_{\rm A/H})$, which means that the limiter values for $\eta_{\rm A/H}$ are essentially in units of $2c_{\rm s}^2/\Omega = 2c_{\rm s}\,H$.} Additionally, we enable the Hall effect only after $30\yr$, so as to allow for a less problematic initial relaxation of the magnetic field in our simulation \citep[also see][]{bai2017global}.

\subsubsection{Radiation and thermodynamics} 

Thermodynamic heating cooling terms are subsumed in the energy source and sink term
\begin{equation}
  \Q \equiv - c\rho\,\kP\! \left(\aR T^4\!-\eR\right)\,+\,\Qirr\,+\,\Qpdr\,,
\end{equation}
where $\eR$ is the radiation energy density, $\aR\equiv 4\, \sigma/c$ is the radiation constant, $\sigma$ is the Stefan-Boltzmann constant, $\kP$ is the Planck mean opacity, and $c$ is the speed of light.

The terms $\Qirr$ and $\Qpdr$ describe the heating (and cooling) via stellar irradiation and a very simplified FUV photo chemistry, respectively. For advancing the radiation energy density, $\eR$, in time, we use the approached detailed in \citet{gressel2017} as well as Section~2.3 of \citet{gressel2020}, which essentially comprises stellar irradiation and diffusive re-radiation (including a flux limiter approach combined with a two-stream approximation to address the optical-depth transition in the vertical direction). Note that, for brevity, we have omitted the radiation force in the above momentum equation, which in general is entirely negligible in our setup. This means that radiative effects are limited to the thermodynamic aspects of our problem, not the dynamic ones.


\subsection{Disk model and boundary conditions}\label{sec:disk_model}

The equilibrium structure of our models is described by a power law relationship of the locally isothermal temperature $T$, with the cylindrical radius, $R$, that is, $T(R) = T_0\,(R/R_0)^q$. A similar relation holds for the midplane density, i.e., $\rho_{\rm mid}(R) =  \rho_0\, (R/R_0)^p$, where we have chosen a temperature slope of $q= -0.75$ (resulting in a mildly flared disk height as a function of radius), and a density slope of $p = -1.5$ --- these are the same values as used in \citet{gressel2020}. The initial solutions for the disk equilibrium \citep*[see][]{nelson2013} then become
\begin{eqnarray}
        \rho(\mathbf{r}) & = & \rho_{\rm mid}(R) \exp{\left[\frac{GM_\star}{c_{\rm s}^2(R)} \left( \frac{1}{r}-\frac{1}{R}\right)\right]}\,,
        \label{eq:dens_eq_sol}\\
        \tilde\Omega(\mathbf{r}) & = & \Omega_{\rm K}(R)\, \left[\,(p+q)\left(
          \frac{H}{R}\right)^2+(1+q) - q\frac{R}{r}\,\right]^{\nicefrac{1}{2}}
        \,,
        \label{eq:Omega_sol}
\end{eqnarray}
where $\Omega_{\rm K}(R) = \sqrt{GM}\!_\star\, R^{-\nicefrac{3}{2}}$ is the well-known Keplerian angular velocity, and $c_{\rm s}^2(R) = c_{\rm s0}^2(R/R_0)^q$ is the sound speed squared. The overall thickness of the disk is controlled by the parameter $H(R_0) = c_{\rm s0}/\Omega_K(R_0)$, and we define $H(R_0) = 0.05\,R_0$, which fixes the overall pressure scale height of the initial gas disc.\footnote{Note that compared with \citet{martel_lesur2022}, our disk is slightly thinner and colder.} The initial density profile was scaled to $\Sigma_0 = 340\,{\rm g\,cm^{-2}}$ at $R_0 = 1\au$, and the aspect ratio is $H(R) = 0.05(R/R_0)^{1/4}$, the mass of the central star $M_\star = 1 M_{\odot}$, the X-ray luminosity of the star is $L_{\rm X} = 10^{29}\,{\rm erg\,s^{-1}}$, and the FUV luminosity, $L_{\rm FUV}=0.2\,L_{\odot}$, is set to be equivalent \citep[see eqn.~(6) in][]{gorti2009} to a mass accretion rate onto the star of $10^{-8}\Msunyr$. We furthermore adopt a mean molecular weight of $\mu=2.4$ and an adiabatic index $\gamma =1.4$, representing a diatomic mixture of molecular hydrogen and helium gas.

In order to mimic a central cavity in our disk, we multiply the profile $\rho_{\rm mid}$ with a function $f(R)$:
\begin{equation}
  f(R) = a_{\rm tr} + \frac{1 - a_{\rm tr}}{2}\;
  \Big[\; 1+\tanh{w_{\rm tr}(R-R_{\rm tr})} \;\Big]\,,
\end{equation}
with a contrast parameter, $a_{\rm tr} = \rhogd $ in the midplane, $R_{\rm tr}$ is the transition radius, where the cavity ends and the disk begins, and $w_{\rm tr}$ controls the steepness of the cavity edge. For all models we have set $a_{\rm tr} = 10^{-5}$, $w_{\rm tr} = 2$, and $R_{\rm tr} = 8$.

We here use the same boundary conditions as in our previous work \citep{sarafidou2024} -- for details we refer the reader to section~3.2.1 of \citet{gressel2020}. At radial boundaries, we apply a standard outflow condition for the momentum density, and extrapolate the mass density with the power-law slope of the equilibrium model. Moreover, standard wave-killing zones are implemented at the inner and outer domain boundary. For the magnetic field components parallel to the domain boundary, we generally enforce vanishing gradients. The remaining perpendicular component is in turn computed from the ${\rm div}(B)=0$ constraint. The outer radial boundary is an exception to this; here we set the azimuthal components of both the field and current to zero. Finally, the axial boundary simply enforces the demanded axisymmetry of our simulations.

\subsection{Integration and averages}\label{sec:integration_and_averages}

Vertical integration is done along a cylindrical height inside the disk,

\begin{equation}
\overline{Q}(R, t) = \int_{-z_d}^{+z_d} Q(R, z, t)dz
\end{equation}
where the integration limits $\pm z_{\rm d}$ are defined as the height where $\betap = 0.1$, a value indicating that there is significant magnetic energy to launch a wind \citep{pascucci2023}.

All models were run for approximately $700$ years, ensuring enough orbital cycles at the transition radius to analyze the cavity dynamics, while taking into account the computational cost (we address the run-time of our simulations in Appendix \ref{app:cavity}). 

Time averaging,

\begin{equation}
\left<Q(R, z)\right> = \frac{1}{T}\int_{t_0}^{t_0+T} Q(R, z, t)dt
\end{equation}
when mentioned, has been done from $450$ to $700$ years, unless stated otherwise.

\subsection{Simulation setup}\label{sec:sim. setup}

The only existing MHD simulations of TDs by \citet{martel_lesur2022} were restricted to the case of AD and OR. Apart from our more detailed thermodynamic and diffusivity treatment, we improve over this work by additionally including the Hall effect (HE) in our simulations. We show the profiles of the non ideal effects in Appendix \ref{app:nonideal}. 

To compare the influence of the HE, we have performed three runs with and without the HE, but with otherwise identical input parameters: besides the disk's power laws being the same, the three runs have also been initialized with the same vertical magnetic field of $\betap = 10^4$ at the disks midplane. The non ideal MHD effects were incorporated as follows: We have two runs that include all three effects (OR,AD, and HE), one with $\mathbf{B}\cdot\hat{\mathbf{z}}<0$ (Hall anti-aligned, \haa) and the other with $\mathbf{B}\cdot\hat{\mathbf{z}}>0$ (Hall aligned, \ha); the third run includes only OR and AD (Hall free, \hf) -- these effects are both invariant under reflection of the field direction, such that it suffices to study one of the field configurations in that case. The mutual orientation of the magnetic field with the rotation axis becomes significant since the HE (in the presence of a vertical field) amounts to a rotation of the horizontal field components in the presence of a horizontal current. Because the sense of rotation depends on the field orientation, this can affect the overall dynamics of the flow.


\begin{figure*}
        \centering
        \includegraphics[width=\linewidth]{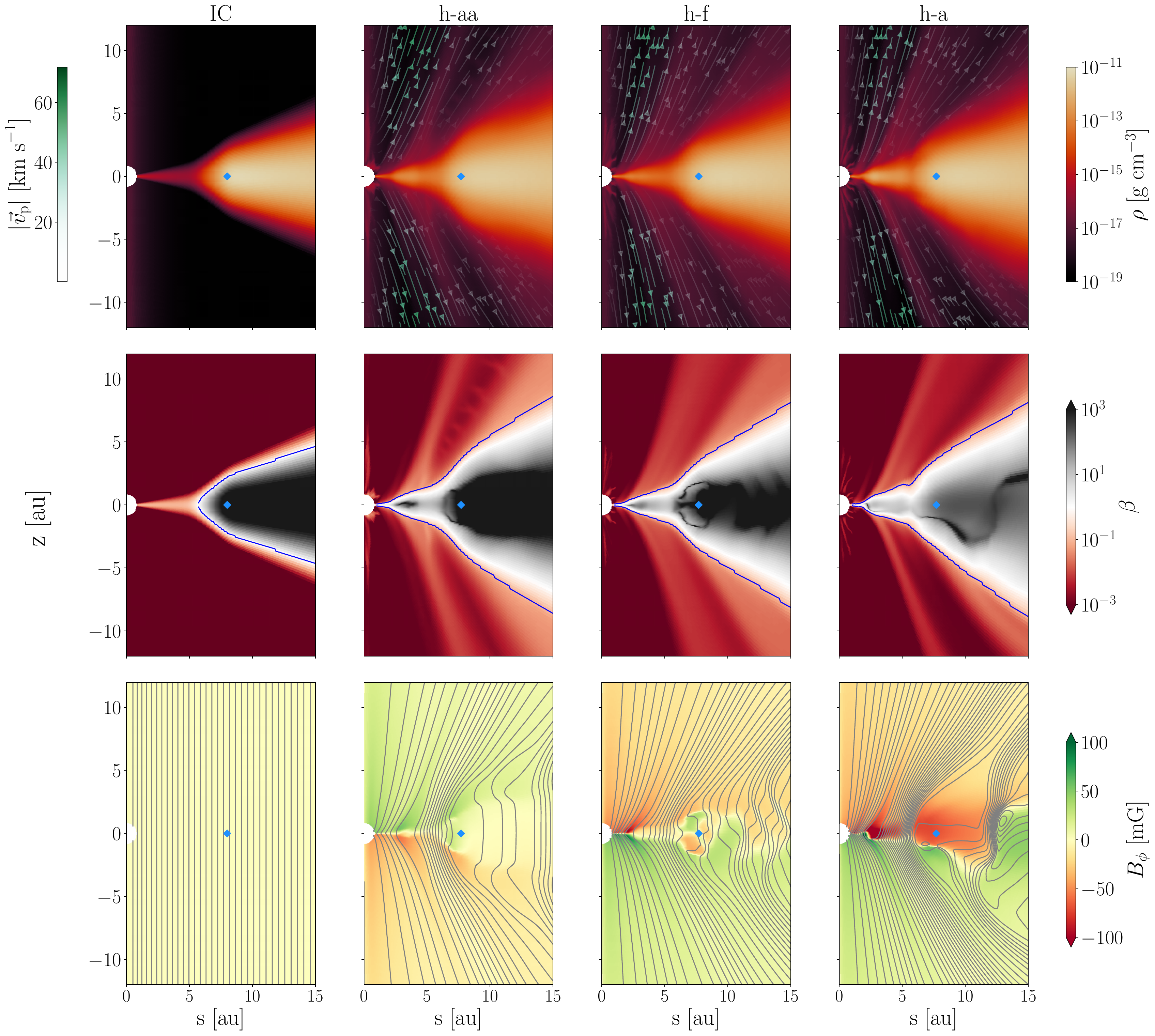}
        \caption{Time-averaged snapshots of the gas density (top row), plasma parameter $\beta$ (middle row), and toroidal magnetic field (bottom row), for: the initial model (first column), as well as the three runs with Hall anti-aligned (\haa, second column), Hall free (\hf, third column), and Hall aligned (\ha, fourth column). The streamlines in the first row indicate the poloidal velocity vectors. The blue lines in the middle row are placed where $\beta = 0.1$ indicating the vertical integration limits $\pm z_{\rm d}$. Finally, the contour lines in the bottom-row panels highlight the poloidal magnetic field lines. The initial radial location of the cavity edge is marked by a blue diamond in all panels.}
        \label{fig:tav_rho_B_b}
\end{figure*}


\section{Results} \label{sec:results}



\subsection{General observations}

In the following, we describe some broad characteristics of our simulations. Looking at the density snapshots in the top row of Figure~\ref{fig:tav_rho_B_b}, we can see that, over the duration of the simulation, the overall extent of the cavity remains roughly stationary for all three runs. While the density contrast with the outer disk is somewhat reduced (compared with the initial setup, shown in the leftmost panel), the cavity remains clearly discernible. Notably, MHD winds originate both in the gap region and in the inner part of the outer disk (at $r\simeq 9 \au$)\footnote{The classical MHD invariants are conserved along wind field lines}. It is important to note that, in comparison to \citet{sarafidou2024}, we do not enforce reflection symmetry with respect to the midplane. As a result, all our models develop a slight top--bottom asymmetry, especially in the case of the \ha, where stronger horizontal fields are produced.

 In the middle row of Fig.~\ref{fig:tav_rho_B_b}, we plot the evolved magnetization (again in terms of $\betap$) for the three cases. We can see that the evolved models features distinct levels of magnetization depending on the assumed microphysics and field orientation: In the case of the \haa, the disk remains close to the initial $\betap$, with $\betap\geq 10^{3}$ in the midplane. Contrary to this, in the case of the \ha, the disk has become significantly magnetized in the most part, with $\betap\sim 10^{2}$ (in the midplane). The central cavity follows a similar pattern, with \ha being the most magnetized, while for the \haa we note a small demagnetized area -- this issue will be discussed in more in detail in Sec.~\ref{sec:rings}, below.

In the bottom row of Fig~\ref{fig:tav_rho_B_b}, we plot the toroidal magnetic field, $\Bph$. Poloidal field lines are over-plotted in terms of contour lines of $R A_{\phi}$, where $A_{\phi}$ the toroidal component of the vector potential \citep[that we evolve alongside the magnetic field variable; see][for details]{gressel2020}. In all three runs, we see symmetry breaking to various degrees. By eye, the \haa appears to retain symmetry better than the other two cases. We moreover see, at the same time, areas devoid of field lines, as well as featuring field line accumulations. Additionally we note that in the \ha case, we have the strongest $\Bph$, while for the \haa model we have the weakest. This is linked to the difference in the disk magnetization and is because of the general tendency of the Hall effect and its mutual interplay with the Keplerian shear.

In the \ha case, there will be an enhancement of the buckling of the field lines, while on the \haa any curvature of the field lines will be severely suppressed \citep{wardle_salmeron2012}. In our models, the magnetic field lines are threading the MHD wind, thus bending and introducing a radial component to the magnetic field. This then gets sheared generating a toroidal magnetic field \citep[see also][]{bai2017hall,sarafidou2024}.

\begin{figure}
        \centering
        \includegraphics[width=\linewidth]{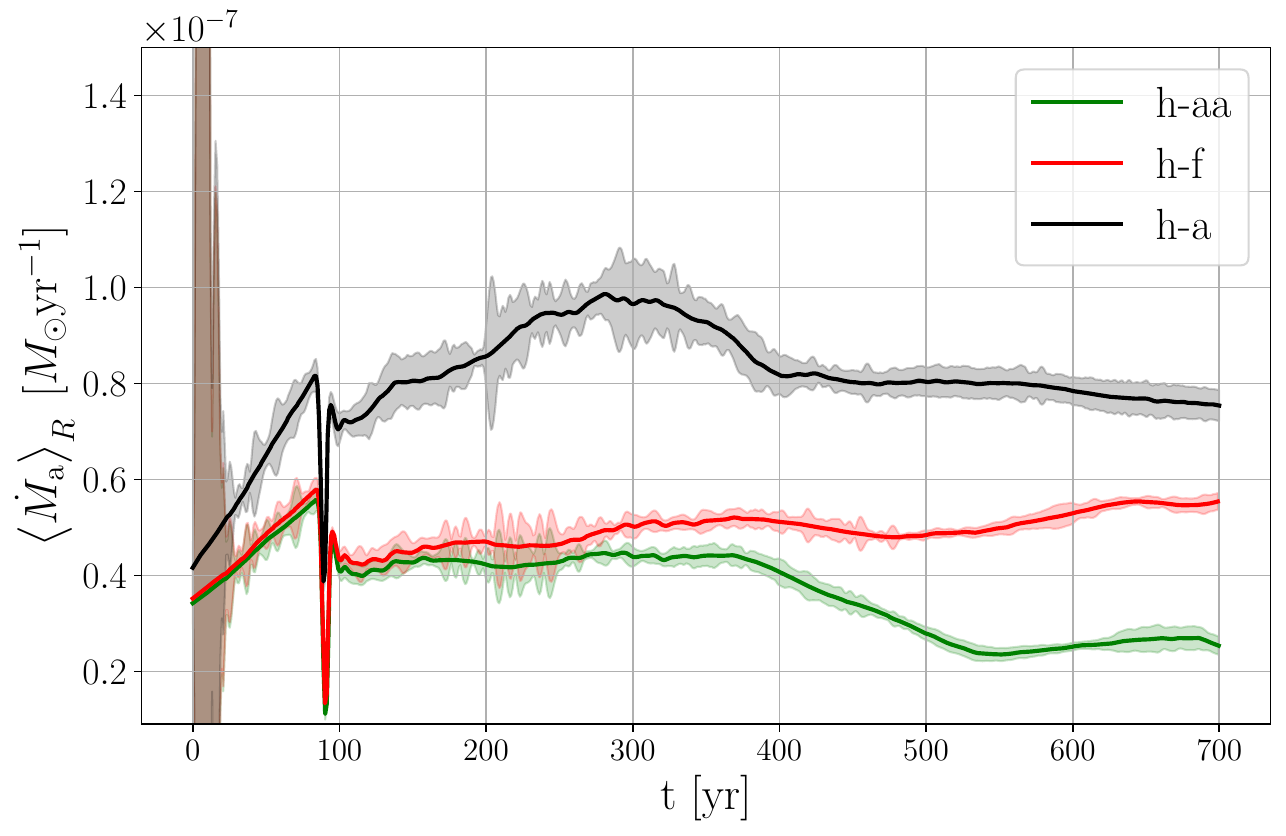}
        \caption{Radially averaged accretion rates through the cavity region over time. Shaded areas indicate uncertainties due to the vertical variability.}
        \label{fig:Ma_vs_t}
\end{figure}


\subsection{Accretion rates in the low-density region}\label{sec:Ma_in_gap}

A central question that we aim to address with our simulations is whether the central cavity can maintain the overall accretion process onto the star, or whether it acts to isolate the outer disk instead. In Fig.~\ref{fig:Ma_vs_t}, we accordingly compare the three models in terms of their mass throughput in the central cavity region over time; the radial accretion rate is simply defined as:
\begin{equation}\label{eq:Ma}
  \Mdota = 2\pi R\overline{\rho \vR}
\end{equation}
where we average over the radial extent $R\in[1, 7]\au$.

As is immediately noticeable, we obtain the highest accretion rate for the \ha case. The \hf and \haa have almost the same accretion rate in the first $200 yr$. After that, though, the \haa produces a somewhat lower $\Mdota$. More specifically, the accretion rates for the \haa, \hf and \ha are: $(2.63\pm 0.31)\ee{-8}$, $(5.13 \pm 0.41)\ee{-8} $, and $(7.91\pm 0.22)\ee{-8} \Msunyr$, respectively. As has been discussed before \citep[see][]{bai2017hall,sarafidou2024}, the fact that the \ha has the highest overall mass accretion rate is a direct result of the aligned Hall effect: the increased radial and toroidal magnetic components lead to a higher overall Maxwell stress, and thus a higher accretion rate.


\begin{figure}
        \centering
        \includegraphics[width=0.7\linewidth]{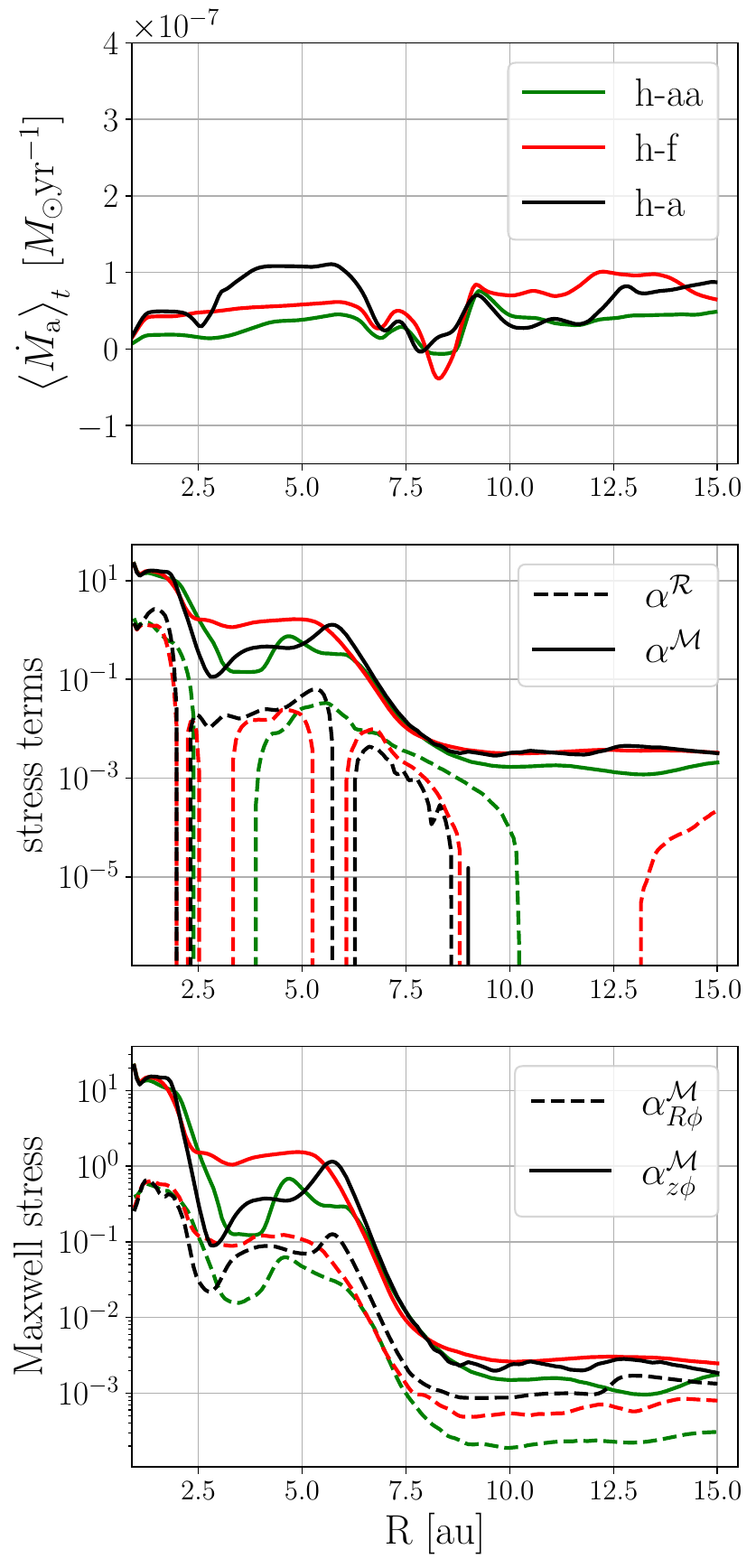}
        \caption{Time-averaged accretion rates and stress terms versus radius for the three runs. In the first panel, we show the measured accretion rates, in the middle panel we show the Reynolds (dashed) vs. the Maxwell stress (solid line), normalized by the vertically integrated pressure. In the bottom panel, we decompose the normalized Maxwell stress into its radial (dashed) and surface (solid line) components.}
        \label{fig:all_accretion_rate}
\end{figure}


For a more detailed analysis of the accretion rate of the disk, we turn to the equation of angular momentum conservation, in its vertically integrated form:
\begin{eqnarray}\label{eq:ang_mom_cons}
  \Mdota = \frac{2\pi R}{\nicefrac{\partial}{\partial_R}\,(\tilde\Omega R^2)} & & \Bigg(\;\frac{1}{R}\frac{\partial}{\partial R} R^2
  \left[\; \overline{\rho\vph'\vR}- \frac{\overline{\Bph\BR}}{\mu_0} \;\right] \Bigg.
  \nonumber\\
  & & + \left.
  R\,\Bigg. \Big(\, \rho\vph'\vz - \frac{\Bph\Bz}{\mu_0} \,\Big)\,
  \right|_{-z_d}^{+z_d}
  \;\Bigg)\,,
\end{eqnarray}
where the mass accretion rate, $\Mdota$, is denoted $-2\pi R\,\overline{\rho\vR}$. In that expression, we have $\R = \overline{\rho\vph'\vR} + \big.R\rho\vph'\vz\big|_{-z_d}^{+z_d}$ as the contribution from the (horizontal and vertical) Reynolds stress, and, moreover, $\M = -\overline{\Bph\BR}/\mu_0 \left.-R\,\Bph\Bz/\mu_0\right|_{-z_d}^{+z_d}$ from the relevant (horizontal and vertical) Maxwell stress components, respectively. We normalize these quantities by the vertically integrated pressure $\overline{p}$ noting them as $\alpha^{\mathcal{R}}_{R\phi} = \mathcal{R}_{R\phi}/\overline{p}$ and so on. Finally, $\vph' = \vph - \tilde{\Omega}R$ defines deviations from equilibrium.


\begin{figure}
        \centering
        \includegraphics[width=0.75\linewidth]{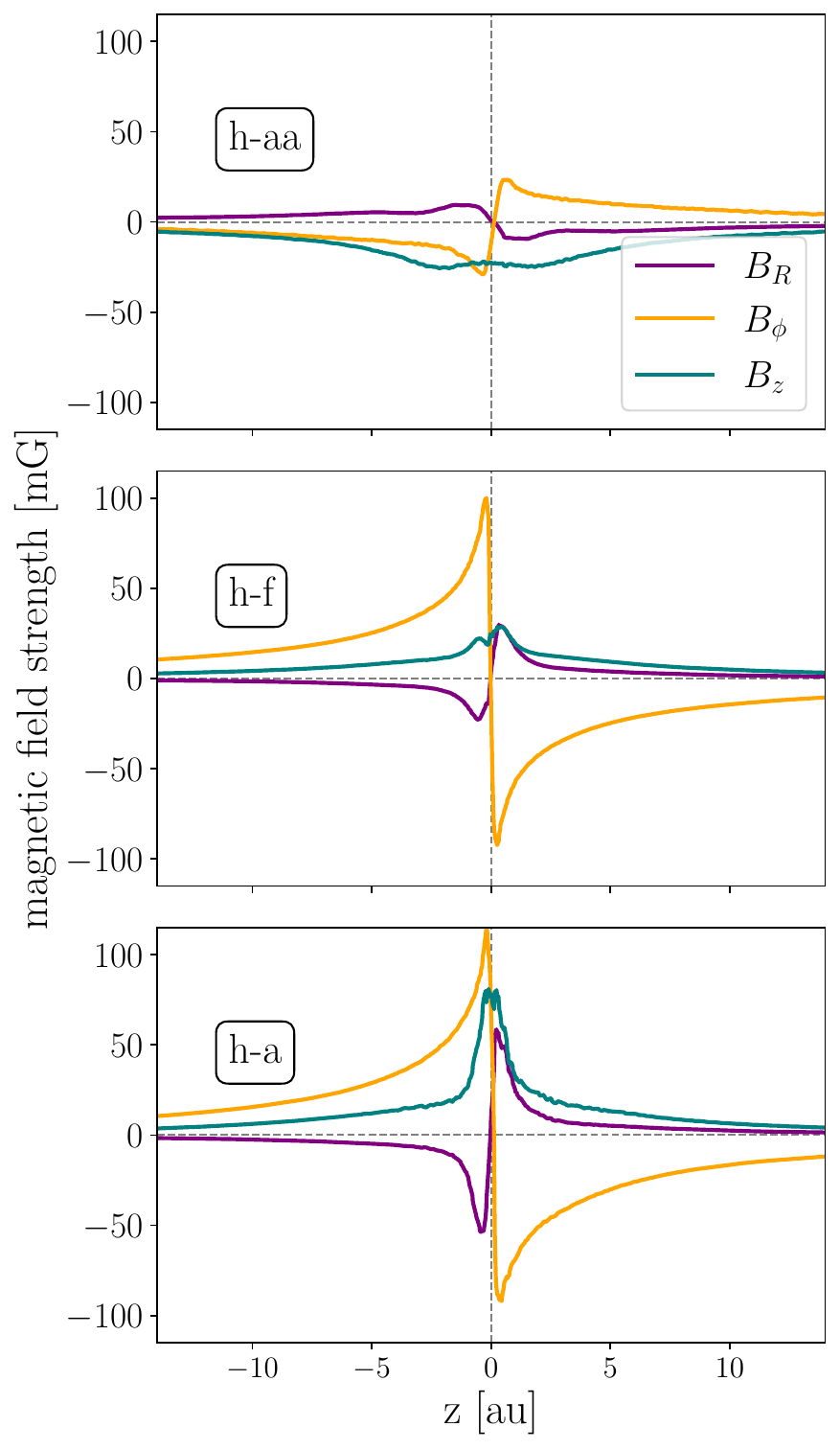}
        \caption{Vertical profiles of the magnetic field components along a poloidal field line (each highlighted by a thick blue line in the bottom panels of Fig.~\ref{fig:tav_rho_B_b}, above) threading the cavity.}
        \label{fig:B_profiles}
\end{figure}


In the top panel of Fig.~\ref{fig:all_accretion_rate}, we show the time-averaged accretion rate over radius for the three runs, calculated from eq. (\ref{eq:Ma}). Apart from the ``dip'' near the transition radius\footnote{We discuss this modulation of the accretion rate in the Appendix.}, in all cases the accretion rates of the cavity and the outer disk are comparable. Looking specifically in the cavity, In the \hf case, the accretion rate is relatively constant, which to a lesser degree is also true for the \haa run, where we mainly see two distinct levels of accretion for the inner and outer half of the region of interest, respectively, and a shallow ramp in-between. In contrast, for the \ha case, we observe a marked dip in the accretion rate at around $r\sim 2.5 - 3\au$ -- and otherwise a similar gradual step at $R\simeq 4\au$ as for the \haa model. The sudden drop in the accretion rate provides an indication that we might observe a substantial level of local mass accumulation just inside this radius, something that we subsequently discuss in detail in Sec.~\ref{sec:rings}. In all cases, the accretion rate stems predominantly from the MHD wind -- see the middle panel of Fig.~\ref{fig:all_accretion_rate}. We see specifically that the surface term of the Maxwell stress is dominant compared to the in-plane stress (as shown in the bottom panel of the figure).

We highlight the specific geometry of the MHD winds in Figure~\ref{fig:B_profiles}, where we plot the magnetic field components (in cylindrical coordinates) for the three runs (top to bottom panels). For all the cases, the horizontal (purple/yellow) and vertical (green) components are roughly comparable in magnitude with each other. This has previously been highlighted in the work of \citet{wang2017}. As they mention, if the magnetic field components are comparable with each other, then the wind inherently has a simple geometric advantage in driving accretion compared to radial (``viscous'') stress. This is because the surface stress acts on a much larger surface area (i.e., $\sim\pi R^2$) compared to the radial stress (i.e., $\sim 2\pi R\, H$). Their analysis further suggests that in such scenarios the accretion velocity is expected to be comparable to the speed of sound. This is further supported when looking at the accretion rates of the models of $\Mdota\sim 3$--$8\ee{-8}\Msun \mathrm{yr}^{-1}$ -- values that are well within the range of expected accretion rates for full disks. Hence if the rates are comparable between full disks and transition disks but the surface density drops by at least two orders of magnitude, then the accretion velocity should be significantly increased.


\begin{figure*}
        \centering
        \includegraphics[width=\linewidth]{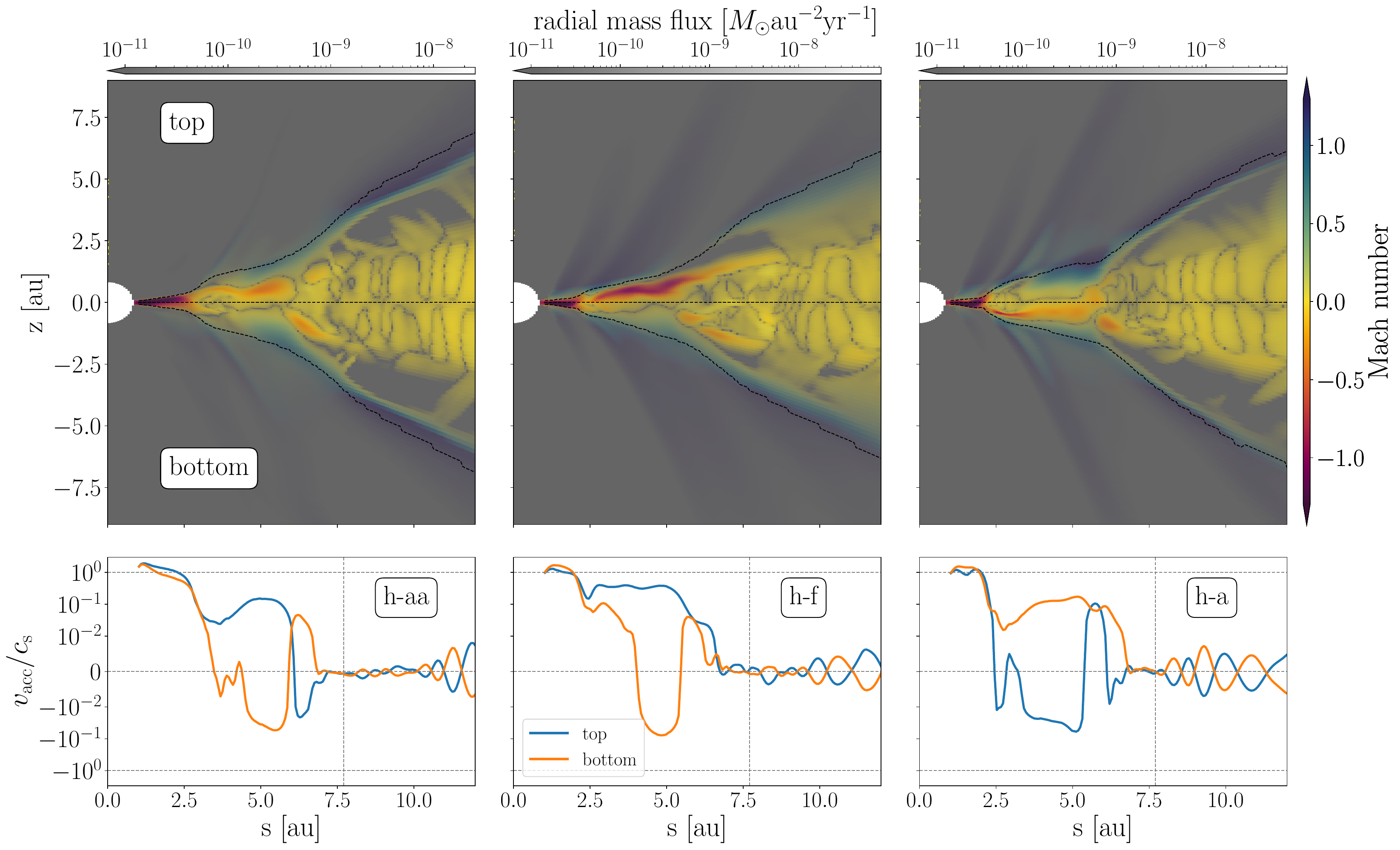}
        \caption{Time-averaged snapshot of the flow Mach number over $100 \rm yr$ (upper panels) and normalized accretion velocity profiles (lower panels) for the three runs. On the snapshots we are overlaying the radial mass flux using a gray alpha gradient colormap. The black dashed lines indicate the disk midplane as well as the disk surface.}
        \label{fig:radial_mass_flux}
\end{figure*}


We distinctly showcase the characteristic feature of near-sonic accretion in Figure~\ref{fig:radial_mass_flux}. Looking at the accretion velocity profiles, we see that there is a clear transition from subsonic flow ($v_{\rm acc}/c_{\rm s} \le 10^{-2}$) in the outer disk to near-sonic flow ($v_{\rm acc}/c_{\rm s} \geq 10^{-1}$) in the central cavity -- where specifically for $R<2.5\au$ we observe transonic accretion for all three runs.

Apart from $R<2.5\au$, where there are non-negligible in-plane Maxwell (and indeed Reynolds) stresses --see the middle and bottom panels in Fig.~\ref{fig:all_accretion_rate}-- accretion actually happens away from the disk midplane. This is a consequence of two fundamental aspects of our simulations: ~a) unlike in classic wind models \citep{blandford1982}, the $\betap=1$ line is at several scale heights above the disk, and ~b) the bending of field lines is suppressed by OR in the body of the disk. Regarding this topological feature of the accretion flow, we highlight that in all three runs we see that a streamer originates from the inner edge of the outer disk (at $R\sim 8\au$) and then traverses the low-density cavity preferring either top or bottom ($\theta:0$--$\pi/2$ or $\theta:\pi/2$--$\pi$), respectively. At the opposite location, an outward mass flux is typically observed -- although not nearly as strong as the corresponding accretion flow through the other disk half. This can be seen in the bottom row of Fig.~\ref{fig:radial_mass_flux} -- note the (signed) logarithmic y-axis in that plot. It is worth mentioning that there seems to be no distinction in the accretion velocity profile between the two different configurations of the HE. Moreover, the position (top/bottom) of the inward streamer is not fixed in time; it varies whenever the current sheet flips its orientation due to the non-vanishing toroidal magnetic field in the disk midplane.


\subsection{Emergent structures within the cavity}\label{sec:rings}


\begin{figure}
        \centering
        \includegraphics[width=\linewidth]{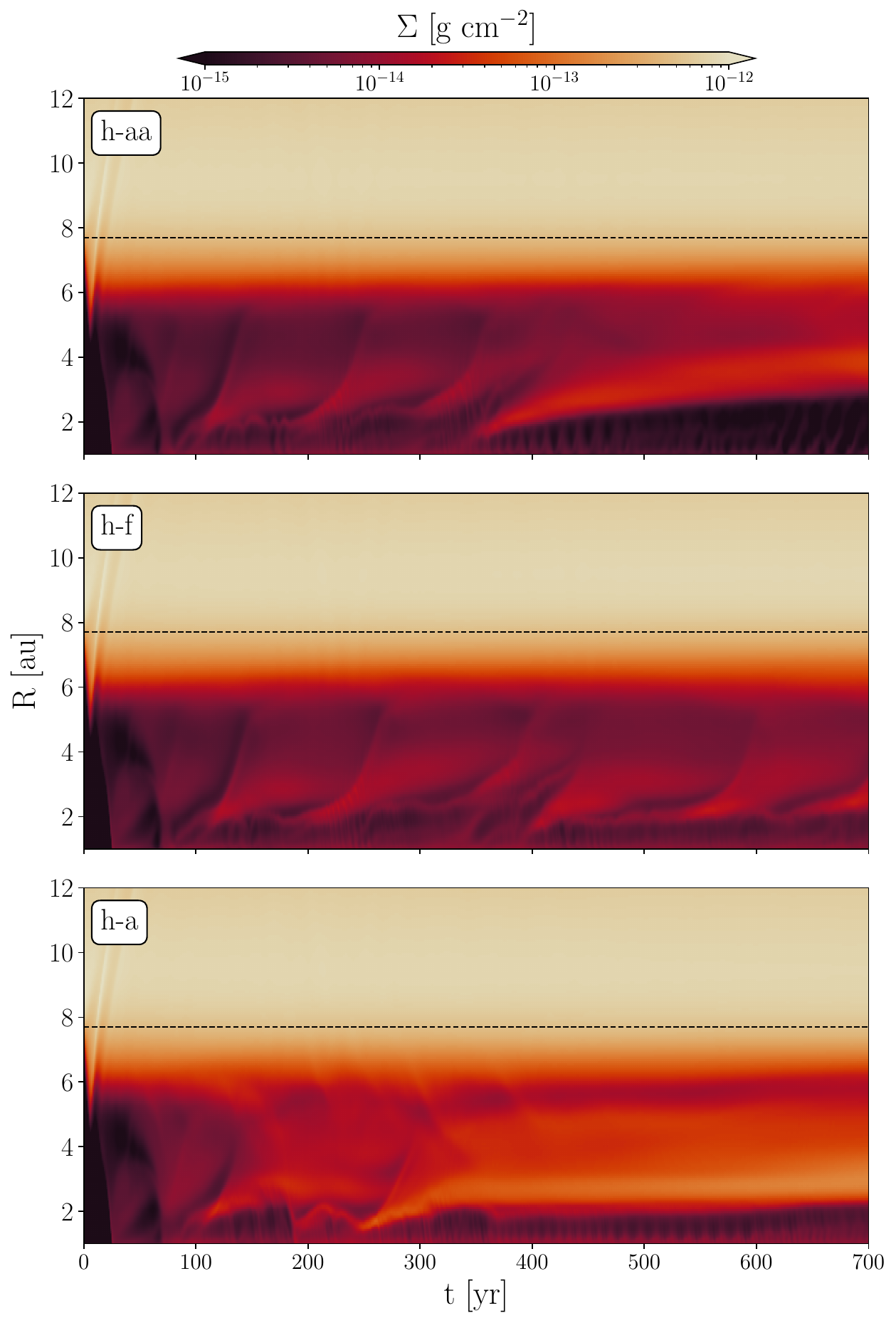}
        \caption{Space-time plot of the surface density for the three runs (from top to bottom: \haa, \hf, \ha)}
        \label{fig:rho_vint}
\end{figure}


\begin{figure}
        \centering
        \includegraphics[width=0.9\linewidth]{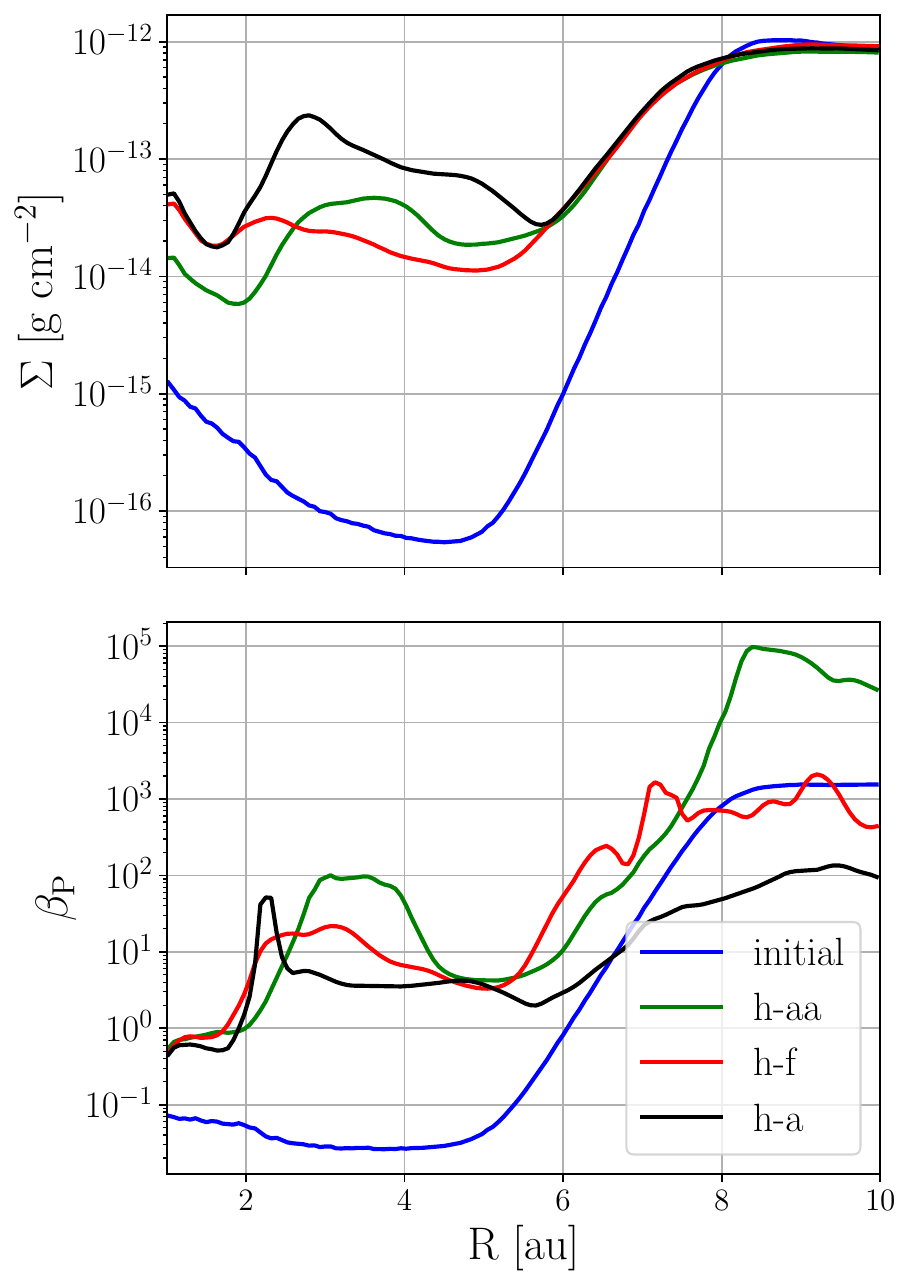}
        \caption{Comparison of the vertically averaged surface density (top) and plasma parameter $\betap$ (bottom panel) over radius for the three runs. The initial profile (blue) is common for all models, the other lines are time averaged.}
        \label{fig:rho_beta_mid_in_vs_fin}
\end{figure}


The time-averaged density snapshots of Figure~\ref{fig:tav_rho_B_b} gave a clear indication that -- at least overall-- the density contrast between the inner cavity and the outer disk dynamically persists in all three runs. This is by no means a given. For a closer look, we provide the space-time plot of the vertically integrated density in Figure~\ref{fig:rho_vint}.

The space-time plot reveals that the gas in the cavity is quite dynamic, with burst of matter appearing (especially in the \hf case  -- cf. the middle panel of Fig.~\ref{fig:tav_rho_B_b}). We also note that in all cases the density of the cavity increased: It was initialized with $a_{\rm tr} = 10^{-5}$, but after $\sim 50yr$ it was partly filled-up. For the \hf case, the cavity contrast, $a_{\rm tr}$, remains at a value $a_{\rm tr} \simeq 10^{-3}$ -- and this sits well within the range of observational estimates for the amount of gas within the gap \citep{vanderMarel2018}. For the two HE cases, though, a different behavior is observed: gas accumulates near the center of the cavity, forming a ring-like structure -- this notion is strengthened by Fig.~\ref{fig:rho_beta_mid_in_vs_fin}, where we clearly see that the two HE runs display broad peaks in their radial profiles. The one formed in the \haa case is the smallest, spanning from $(2.5$--$5.5)\au$. The ring in the \ha case is slightly larger, from $(2.0$--$5.5)\au$.


\begin{figure}
        \centering
        \includegraphics[width=0.9\linewidth]{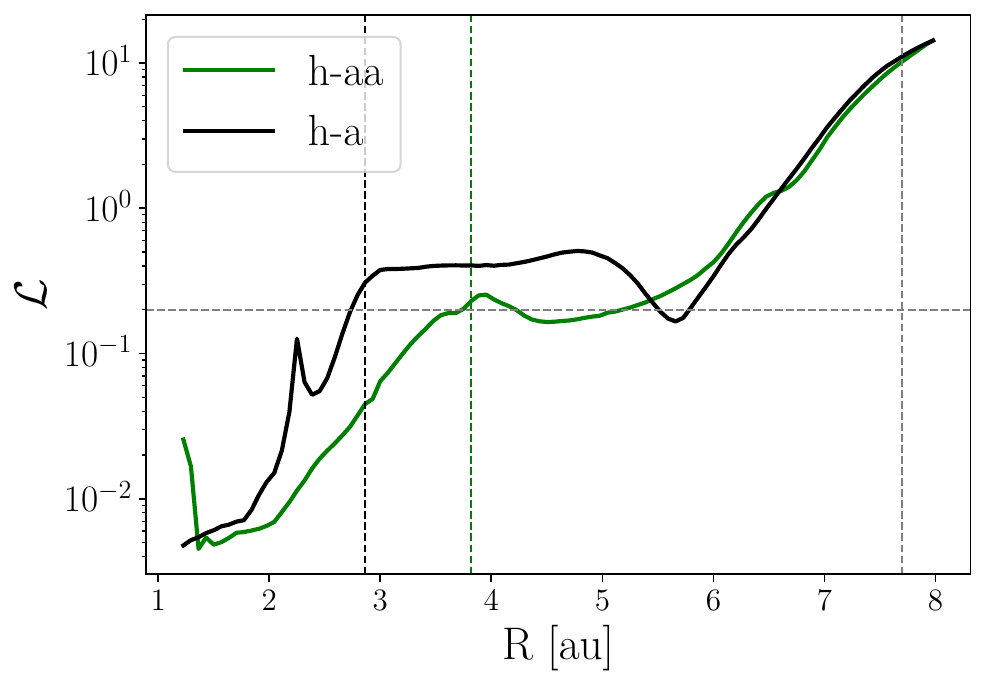}
        \caption{Radial profiles of the $\mathcal{L}$ parameter for the two HE runs. The vertical dashed black and green lines indicate the density maxima of the ring in the \ha and \haa case, respectively. The vertical gray dashed line indicates the outer rim of the cavity. }
        \label{fig:L_hall_runs}
\end{figure}


Since these ring-like structures only appear in the HE runs, we explore the Hall-MHD self organization mechanism that was introduced in \citet{kunz_lesur2013} and further detailed in \citet{bethune2016}. This mechanism requires the system to be in the ``strong'' Hall regime, characterized by the dimensionless parameter $\mathcal{L} = l_{\rm H}/H \gtrsim 0.2$, where $l_{\rm H} = \etaH/v_{\rm A}$ is the ''Hall length'', relating the Hall coefficient $\etaH$ to the Alfv{\'e}n speed $v_{\rm A} = |B|\,/\!\sqrt{\mu_0\rho}$ -- also cf. \citet{kunz_lesur2013}. Moreover, the in-plane component, $\mathcal{M}_{R\phi}$, of the Maxwell stress tensor is characteristically found anticorrelated with the vertical magnetic field. This arises because the Maxwell stress appears in the induction equation through the Hall term. According to \citet{bethune2016}, the Hall-MHD induction equation can be expressed as
\begin{equation}
  \frac{\partial B_z}{\partial t}\simeq
  \Big(\nabla\times\left[\V\times\B\right]\Big)_{z}
  -\, l_{\rm H}\, \frac{\partial }{\partial R^2}B_R B_\phi\,.
\end{equation}
This illustrates that --in the presence of a strong HE-- the curvature of the in-plane Maxwell stress directly influences the induction equation, indicating a strong link between angular momentum transport and vertical magnetic flux: when the second derivative of $\mathcal{M}_{R\phi}$ is negative, vertical magnetic flux is pushed outward. Consequently, the vertical magnetic field $B_z$ should be anticorrelated with $\mathcal{M}_{R\phi}$, where a local maximum in $\mathcal{M}_{R\phi}$ corresponds to a local minimum in $B_z$.


\begin{figure}
        \centering
        \includegraphics[width=0.9\linewidth]{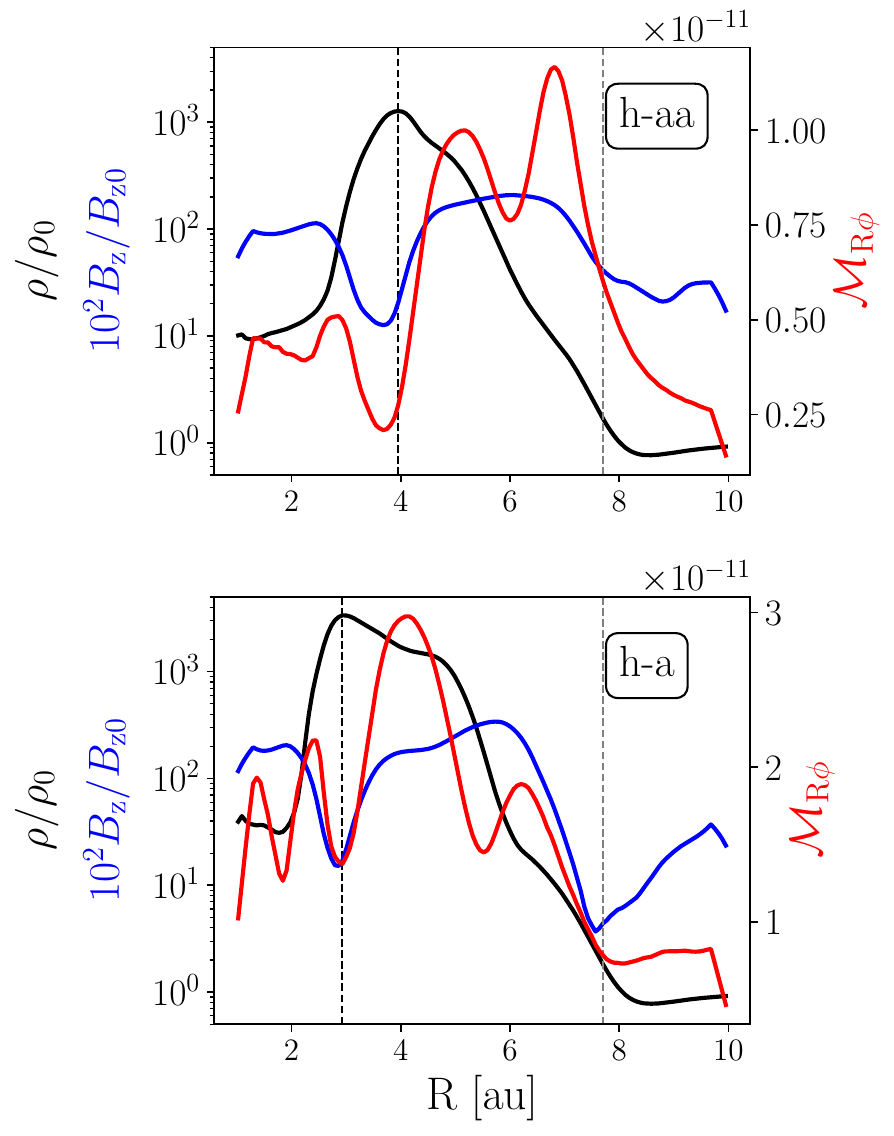}
        \caption{Vertically integrated profiles of the density, vertical magnetic field, and radial Maxwell stress in the cavity. We note the anti-correlation between density and $\Bz$. The black dashed line (on the left) indicates the density maxima, while the gray dashed line (on the right) marks the end of the cavity.}
        \label{fig:rho_Bz_Mrf}
\end{figure}


While in the cavity we are indeed in the strong Hall regime --see Figure~\ref{fig:L_hall_runs}-- the radial Maxwell stress is not anti-correlated with $B_z$. From Figure~\ref{fig:rho_Bz_Mrf}, it in fact looks like they are actually correlated, instead.


\begin{figure}
        \centering
        \includegraphics[width=0.8\linewidth]{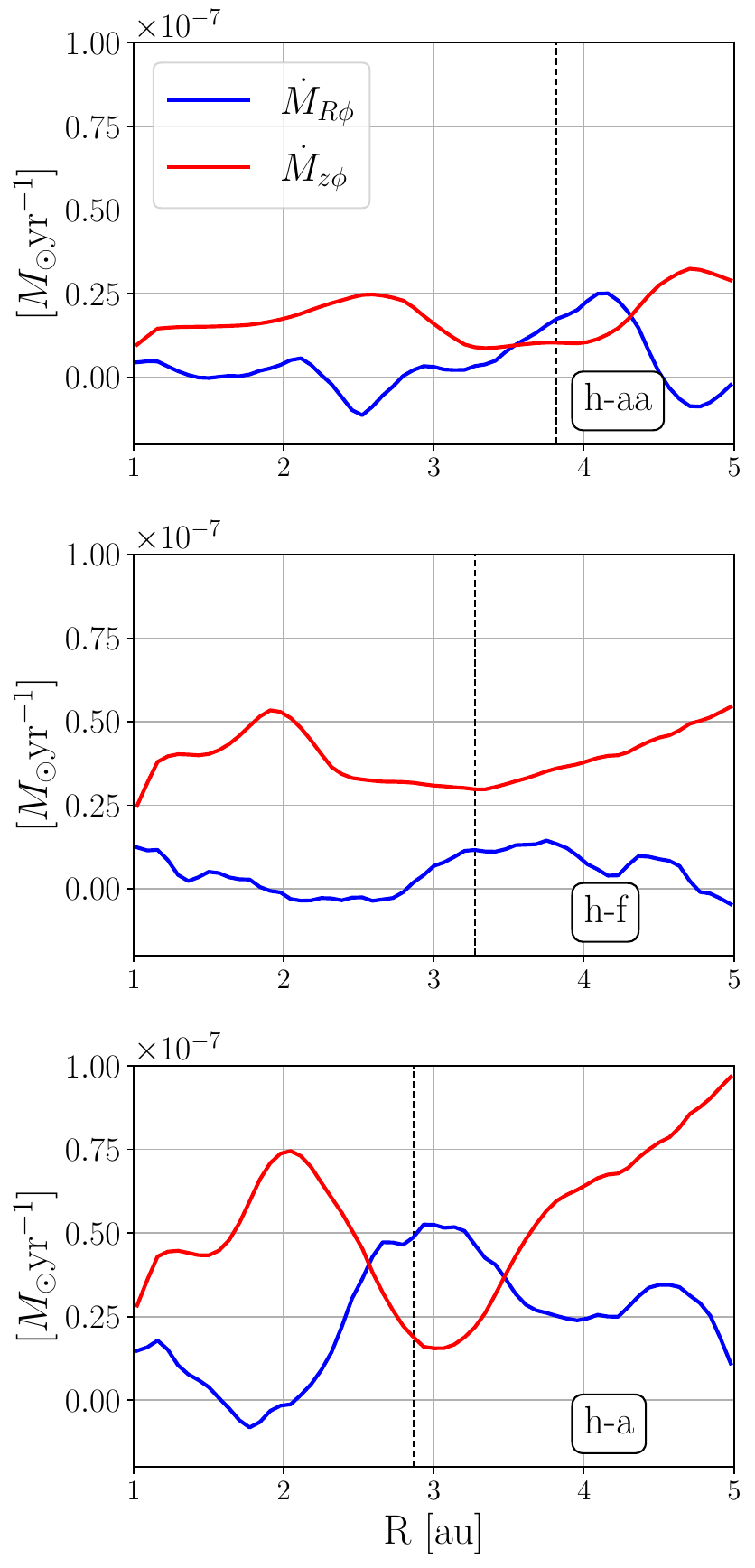}
        \caption{Accretion rate in the cavity due to the wind ($\dot{M}_{z\phi}$) and the radial stress ($\dot{M}_{R\phi}$ - see eq. \ref{eq:ang_mom_cons}) for the three runs.}
        \label{fig:wind_vs_radial}
\end{figure}


The only other significant difference between the Hall and Hall-free runs, which might explain the ring formation in the former but not the latter, is observed when comparing the accretion rates from the wind to those from the radial stress, as shown in Figure~\ref{fig:wind_vs_radial}.

In regions with ring formation, only in the Hall cases do we observe $\dot{M}_{\rm rad} > \dot{M}_{\rm wind}$, indicating that in those areas, more mass is replenished in the gap than is being ejected by the wind. This is remarkably similar to the wind driven instability that was described in  \citet{suriano2017,riols_lesur2019}: In non ideal MHD, as mass is accreted, magnetic flux diffuses away from the region of mass concentration, leading to a decrease in magnetic flux where density accumulates (see the vertically integrated profiles of density and $B_z$ in Fig.~\ref{fig:rho_Bz_Mrf}). This results in a dynamically weakened magnetic field, that in turn increases the accretion timescale in that region. Consequently, this creates an ``accretion trap'' for mass that is transported from larger radii (most apparent in the radial accretion profile of the \ha in the top panel of Fig. \ref{fig:all_accretion_rate}). Magnetic flux is then accumulated in the under-dense regions strengthening the wind and more effectively removing angular momentum. The reason why this is only happening in the Hall runs is currently unclear.


\subsection{Magnetic topology and flux evolution}


\begin{figure}
        \centering
        \includegraphics[width=\linewidth]{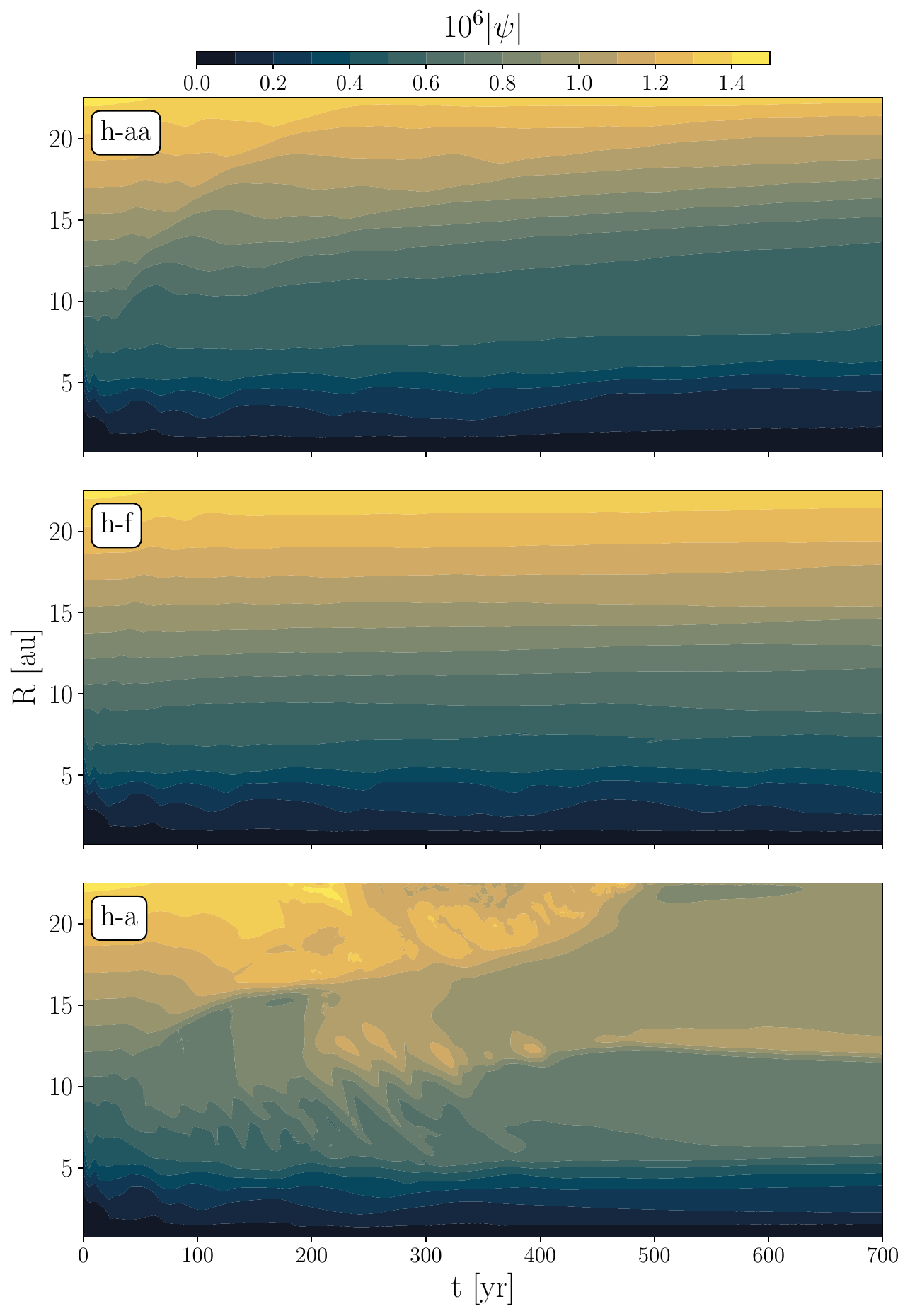}
        \caption{Space-time evolution of the poloidal magnetic flux function $\,\psi(\theta=\pi/2)\,$ for the three runs (\haa, \hf, \ha).}
        \label{fig:mag_flux_spacetime}
\end{figure}


There is a broad consent that the efficiency of any accretion and/or mass ejection mechanism that relies on magnetic forces has to somehow scale with the magnetic field strength. This equally holds for torques from turbulent (laminar) Maxwell stresses, as well as for magnetic tension (pressure gradient) forces in magnetocentrifugal (pressure-driven) disk winds. To obtain a robust handle on disk evolution, one hence needs to understand the topology and evolution of the entrained magnetic flux, that is threading the disk.

In this section we thus want to look at the temporal evolution of the magnetic flux. To do that we utilize the poloidal magnetic flux function
\begin{equation}
  \psi = r\sin(\theta)\,A_\phi\,,
\end{equation}
where the azimuthal component, $A_\phi$, of the magnetic vector potential is evolved alongside the induction equation. During our \nirv simulation runs, the simple equation \citep[also cf. sect.~2.2.1 in][]{gressel2020}
\begin{equation}
  \partial_t A_\phi = - \mathcal{E}_\phi\,,
\end{equation}
is solved, with $\E$ given by eqn.~(\ref{eq:emf}) in Section~\ref{sec:methods}. Because isocontours of the flux function $\Psi$ describe poloidal field lines, we can use it to study the redistribution of magnetic flux. Looking at the space-time plots in Fig.~\ref{fig:mag_flux_spacetime}, we see that for the \hf case (middle panel) and --to a somewhat lesser degree-- for the \haa case (top panel), the flux appears almost stationary over the timescale of our simulation runs. In contrast, for the \ha case, we have a different picture. We observe the flux in the outer disk (i.e., for $R>8\au$) being pushed outward. Moreover, on a secular timescale, a lot of fluctuations emerge near the transition radius. At later times, this transient variability appears to have settled down, though, and gives way to a quasi-stationary state of the field evolution.

To gain a better insight into the efficiency and timescale of the transport, we use the radial motion of the vertical magnetic flux in the disk midplane as a simple proxy. To this end, we define the magnetic flux velocity \citep[see also][]{bai2017hall} relative to the Kepler velocity, $\vK$, that is
\begin{equation}
  \vB = \frac{\E_{\phi}}{\Bz\vK}\,.
\end{equation}


\begin{figure}
        \centering
        \includegraphics[width=0.8\linewidth]{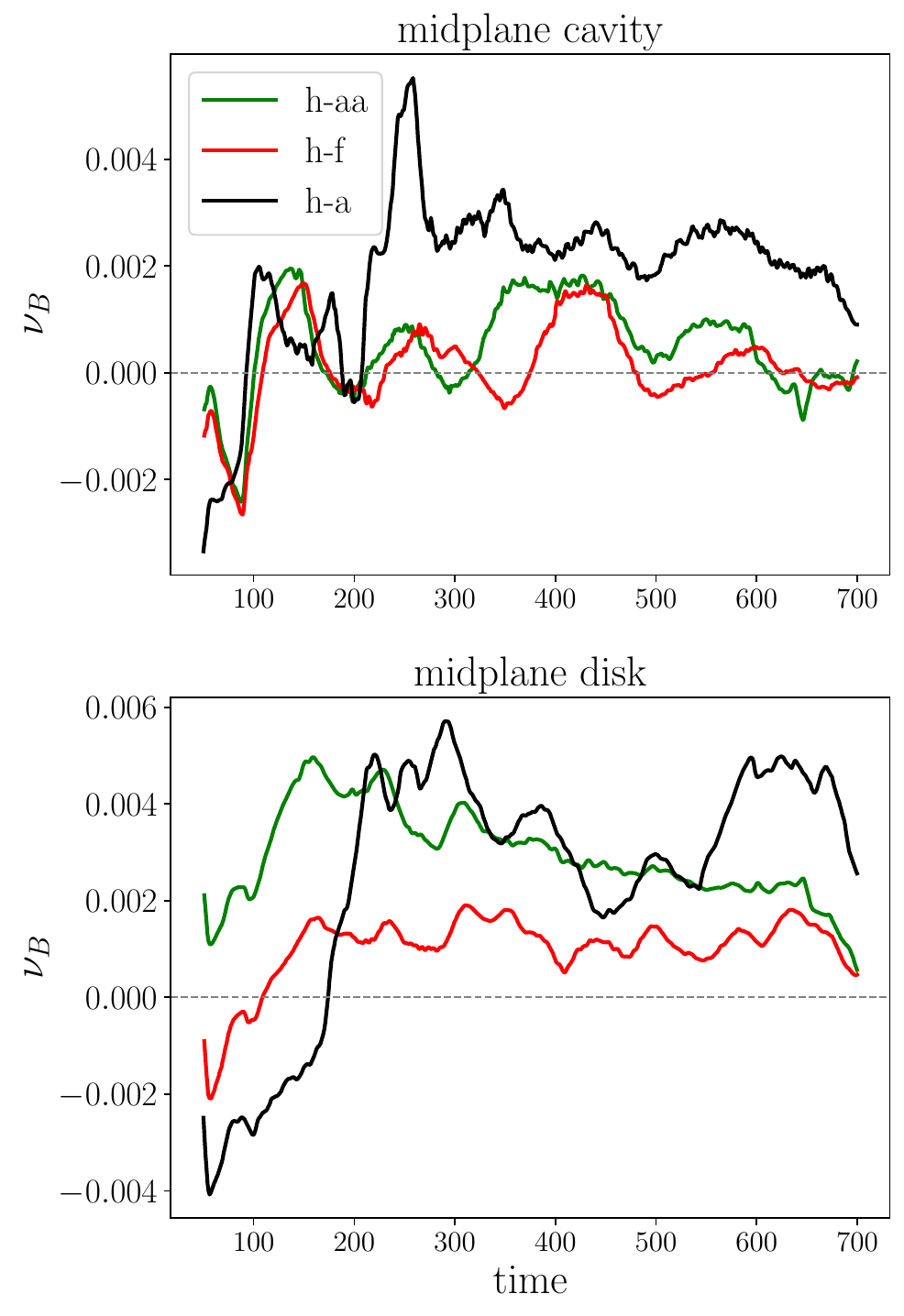}
        \caption{Magnetic flux velocity in the midplane over time.}
        \label{fig:vB_vK_time}
\end{figure}


\begin{table}
        \centering
        \footnotesize
        \begin{tabular}{lcc}
                \hline
                \\[-4pt]
                Simulation  & $\vB$ cavity & $\vB$ outer disk \\[+2pt]
                \hline
                \hline\\[-6pt]
                \haa & $0.7\ee{-3}$  & $2.5\ee{-3}$ \\
                \hf  & $0.2\ee{-3}$  & $1.2\ee{-3}$ \\
                \ha  & $2.3\ee{-3}$  & $3.4\ee{-3}$ \\[+2pt]
                \hline
        \end{tabular}
        \caption{Time-averaged values of $\vB$ in the midplane.}
        \label{tab:vB_values}
\end{table}


In Figure~\ref{fig:vB_vK_time}, we plot $\vB$ for the location of the central cavity (top panel) and, for comparison, also that of the outer regular disk (bottom panel). Comparing the overall behavior of $\vB$ in the cavity versus that in the disk, and focusing on the time interval $t>200\yr$, we note that while in the latter we mainly see outward transport, in the former we observe significant fluctuations. With the exception of the \ha case, $\vB$ is mainly outward but sometimes inward.

Between the three runs, the \ha seems to have the fastest outward magnetic flux transport -- a fact that can also be deduced from the sudden flux evacuation visible in the bottom panel of Fig.~\ref{fig:mag_flux_spacetime}. The average values of $\vB$ are shown in Table~\ref{tab:vB_values}. We see that overall $\vB$ is higher in the disk - in the case of the \haa and \hf the magnetic flux of the cavity is almost unchanged.


\section{Discussion} \label{sec:discussion}


The results presented in the previous section demonstrate that TDs with central ``cavities'' --that is, with a significant depression in the gas surface density-- are robust and long-lived entities. Moreover, their characteristics remain broadly stable when evolving the non ideal MHD equations over hundreds of inner orbits. This is specifically true when focusing on the effects of OR and AD only \citep[also cf.][]{martel_lesur2022}. The picture is somewhat altered when including the HE -- both in its aligned and anti-aligned field configurations. In these cases, we do observe a certain tendency for a radial modulation of various quantities -- such as, for instance, the surface mass density, the vertical magnetic flux, as well as the local accretion rate and the wind mass-loss rate. The overall ability of the disk to accrete material efficiently through the low-surface density region is, however, not affected when including the HE. In the contrary, in the aligned configuration, \ha, the accretion rate even gets boosted by more than 50\%.

Regarding the magnetization of the disk, \citet{martel_lesur2022} concluded that the cavity self regulates to $\betap \sim 1$--$10$. This is in agreement with our Hall-free run. However, for the two HE cases, \haa, and \ha, in the regions where we have the rings, $\betap$ naturally increases, reaching $\betap\simeq 100$. Another difference with \citet{martel_lesur2022} arises in regards to the mass accretion rate within the cavity: in their equivalent Hall-free run it is more than three times larger than ours -- that is, $(\sim 1.4\ee{-7}$ compared to $\sim 4\ee{-8} \Msunyr)$. We believe this discrepancy arises from differences between our respective models. Despite efforts to maintain comparable initial parameters, notable distinctions remain. These include the implementation of the diffusivities of the non ideal effects, the inclusion of thermochemical heating, and an overall thinner disk. Nevertheless, both values fall well within the range of typical inferred accretion rates for various observed TDs \citep[see e.g.,][]{espaillat2014}. This is also true for the accretion rates in the Hall runs. Even though the \ha (\haa) has the highest (lowest) rate, the values range from $(2$--$8)\ee{-8}\Msunyr$ -- this may potentially be responsible for some of the variation in the observational scatter in $\Mdota$. Additionally, in all three runs we observe transonic accretion at small radii. This is something that is expected and, in fact, has been noted in a number of previews works: there is no clear distinction in the accretion rate of a transition disk in comparison with a full disk \citep[e.g.,][]{fang2013,manara2014}). As \citet{wang2017} point out, a steady accretion rate combined with a several magnitude drop in density should naturally lead to (trans)sonic accretion speeds.

Magnetic flux transport is a topic that has been extensively studied \citep[e.g.,][]{bai2017hall,gressel2020,lesur2021,martel_lesur2022}, yet it remains puzzling. While differences in models and simulations across these studies certainly account for some of the scatter in reported values, there is still surprisingly persistent disagreement over the transport rates -- as recently highlighted by \citet{lesur2021}. For instance, \citet{bai2017hall} found that the \haa configuration has the highest magnetic velocity advection $\vB$, while in our case, it is the \ha configuration. \citet{gressel2020} reported higher $\vB$ values for the Hall-free run, whereas \citet{martel_lesur2022} observed inward transport. \citet{lesur2021} suggested that $\vB$ is strongly dependent to the strength of the Ohmic diffusivity - and to a lesser extent to the ambipolar diffusivity. The different implementation and treatment of the diffusivities might indeed be the case for these discrepancies -- or the different evolution of the system in the cases where the setup is otherwise the same \citep[i.e.,][who simulated non-TD PPDs]{gressel2020}.

Similarly puzzling is the emergence of annular rings in our HE simulations. Self-organization into annular rings or so-called zonal flows is something that has been reported consistently in MHD simulations with a net magnetic field over the last decade -- see \citet{riols_lesur2019} for a detailed discussion. However, the mechanisms behind the emergence of these structures seem to be diverse. \citet{bai_stone2014} and \citet{bai2015hallII} found that the structures observed in their models are a robust outcome of the magnetorotational instability (MRI) with a net vertical magnetic flux, and these structures are further amplified by ambipolar diffusion.  \citet{kunz_lesur2013}, \citet{bethune2016}, and \citet{krapp2018} attributed their rings to the HE and the confinement of $B_z$ from the radial Maxwell stress. \citet{bethune2017} deducted that the cause is ambipolar diffusion, acting anti-diffusively locally, due to the opposite signs of $J_{\phi}$ and $J_{\phi,\perp}$; \citet{suriano2018,suriano2019} also attributed ring formation to ambipolar diffusion, though the magnetic structure that it develops in the disk.

In this work, we noticed that in the two HE runs, the accretion rate due to the radial Maxwell stress is actually larger than that from the MHD wind (i.e., $\dot{M}_{R\phi}>\dot{M}_{Rz}$), in the under-dense region before the ring is formed. As has been mentioned before, this is very similar to what was reported in \citet{suriano2017,riols_lesur2019} - however we did not collect convincing evidence as to why this should only happen in the HE runs.


\section{Conclusions} \label{sec:conclusions}


We have performed 2D global axisymmetric R-MHD simulations of TDs, specifically including the HE and a thermochemistry module \citep{gressel2020} for a more realistic temperature treatment\footnote{Even though we do not discuss any issues related to the thermodynamics of the wind launching in the current paper.}, as well as dynamically evolving micro-physical diffusion parameters, $\etaO$, $\etaH$, and $\etaA$.

Our paper highlights results from three runs with the same initial parameters -- the only difference lies in the configuration of the Hall effect: we had the Hall anti-aligned effect (\haa) where the HE acts to suppress any inward curvature of the magnetic field lines in the poloidal plane; the Hall free (only OR and AD, termed \hf); and the Hall aligned, where the HE enhances any concave deformation of the field lines, thus favoring the amplification of laminar fields within the disk. Our main conclusions are as follows:

In all three runs we observe a sustained cavity that retains its size, and an outer disk that behaves like a laminar disk with MHD winds. Overall we observe outward magnetic flux transport, with the transport velocity within the cavity being somewhat lower than that of the outer disk. The cavity features a similar level of magnetization for all runs, with $\betap \sim 1$--$10$. As expected from previous simulations without a cavity \citep{sarafidou2024}, an increasing trend in magnetization is observed in the runs \haa to \hf to \ha.

This increasing order also holds for the measured accretion rates of the cavity, with $\Mdota\sim3\ee{-8}, 5\ee{-8}, 8\ee{-8}$ for the \haa, \hf, \ha case, respectively. This is due to the significantly amplified horizontal magnetic field that is the result of the \ha configuration, that in turn leads to an increased in-plane Maxwell stress. While decomposing the stresses, we find that for all cases, the magnetic Maxwell stress is significantly higher than the purely hydrodynamic Reynolds stress. Specifically, it is the vertical Maxwell stress (i.e., the stress owing to the MHD wind) that contributes the most toward the mass accretion rate. This is a result that we have already seen in full disks and that comes to support the notion that TDs behave like full disks.

A direct result of the similar accretion rates in TDs and full disks is that the gas accretion velocity is expected to be close to sonic or transonic. Indeed, for all runs, the accretion velocity within the cavity is $\sim c_{\rm s}$ for $r\in(2.5, 5.5)\au$ and even $>c_{\rm s}$ for $r<2.5\au$. We also notice that the bulk of the accretion comes from an accumulated radial mass flux (not unlike what has been termed a ``streamer'' in the literature) that originates from the inner radius of the outer disk (at $r\sim7.5\au$) and permeates the cavity above or below the midplane, from a height of $\sim H/R = 0.2$ \citep[cf. also][who speculate about a Rayleigh-Taylor origin of the avoidance of the midplane]{martel_lesur2022}.

At the same time, we observe some level of self organization within the flow, where we have the formation of a ring in the cavity -- but only in the case of the HE runs. We believe that this is a result of the MHD wind removing angular momentum from the gap more efficiently than in regions of high surface density -- even though it is currently unclear why this did not occur in a comparable manner in the Hall free case. The formation of rings in global MHD disks with magnetic winds calls for more detailed study: It has been shown
\cite[see, e.g.,][]{riols_lesur2018,krapp2018} that the radial pressure profile resulting from the sub-Keplerian rotation of the zonal rings can trap dust. As a consequence, such features could well have implications for the accumulation of small solids -- with observationally detectable features, for example, in the dust continuum emission, or more generally, for boosting planet forming mechanisms.


\begin{acknowledgements}
  This work used the \nirv MHD code version 3.8, developed by Udo Ziegler at the Leibniz-Institut f{\"u}r Astrophysik Potsdam (AIP). Computations were performed on the \textsc{taurus} node of the local AIP cluster. We acknowledge support of the DFG (German Research Foundation), Research Unit `Transition discs`, funding ID 325594231. This work was co-funded\,\footnote{Views and opinions expressed are however those of the author(s) only and do not necessarily reflect those of the European Union or the European Research Council. Neither the European Union nor the granting authority can be held responsible for them.} by the European Union (ERC-CoG, \textsc{Epoch-of-Taurus}, No. 101043302).


\end{acknowledgements}


\bibliographystyle{aa} 
\bibliography{ref} 

\begin{appendix}

\section{Assessment of the cavity longevity}\label{app:cavity}

Our simulations may not have run for long enough to unambiguously confirm that they have fully reached a steady state -- that is, with respect to the sustained coexistence of an inner cavity and an outer disk over extended timescales. This limitation stems from the intrinsic complexity of our model and the associated computational cost. However, to address this issue, we conducted a lower-resolution, Hall-free simulation (run over $12,000$ years) to assess whether the cavity can be maintained over these timescales. In the following, we briefly present the main results of this fiducial study.

\begin{figure}[h]
  \centering
  \includegraphics[width=\linewidth]{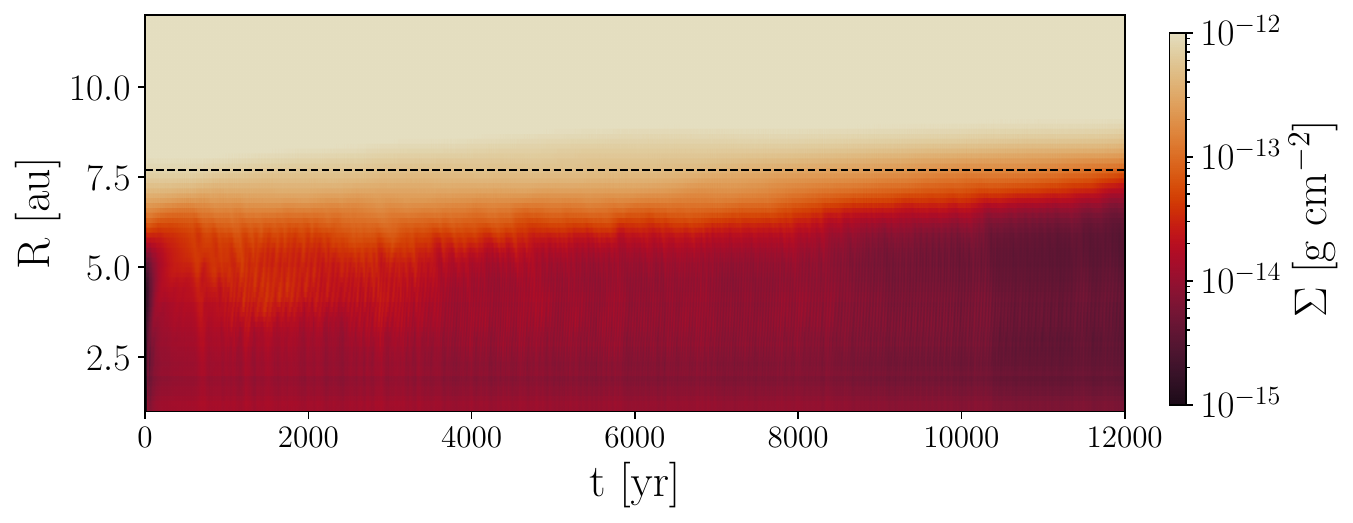}
  \caption{Space-time plot of the vertically averaged density.}
  \label{fig:hf_lowres_spacetime}
\end{figure}

Figure~\ref{fig:hf_lowres_spacetime} presents the space-time plot of the vertically averaged density, which displays an overall steady state despite recurring high-frequency variations in the density. These results are consistent with Fig.~3 of \citet{martel_lesur2022}, who also observe significant density variability within the cavity. Given that their models evolved stably over longer timescales, there is no indication that our models would behave differently.

\begin{figure}[h]
  \centering
  \includegraphics[width=0.9\linewidth]{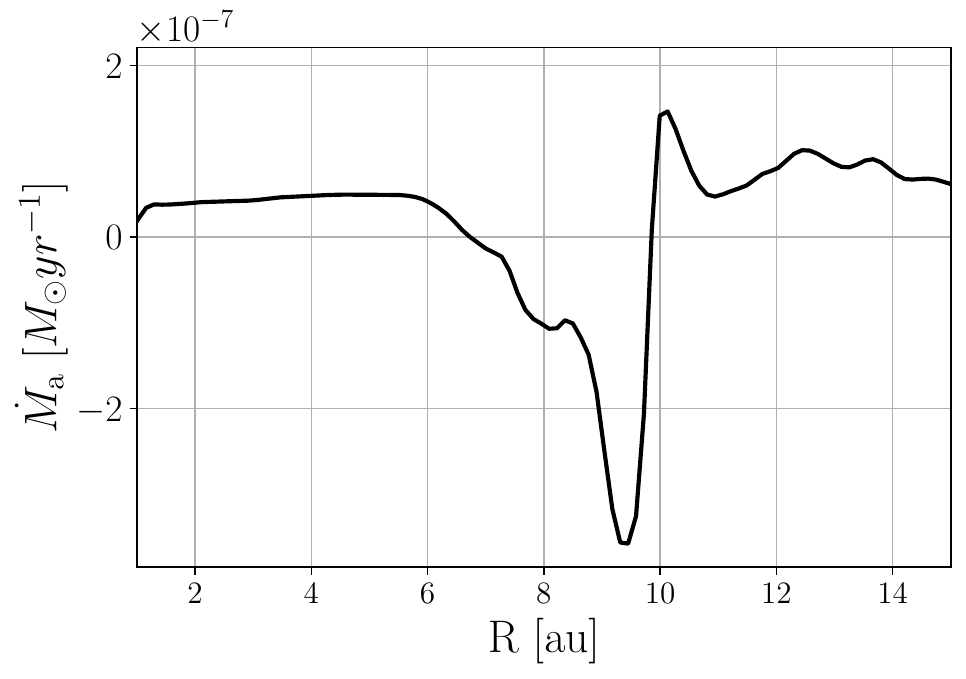}
  \caption{Time-averaged accretion rate.}
  \label{fig:r_accretion_hf_lowres}
\end{figure}

Figure~\ref{fig:r_accretion_hf_lowres} shows the accretion rate as a function of radius, averaged over the last $5,000$ years of that simulation. Notably, the radial profile of the accretion rate shows a dip (with negative values) just inside the original transition radius, while it features a peak outside of that radius. The corresponding Fig.~8 in \citet{martel_lesur2022} shows a similar non constant profile of the accretion rate with radius, depending on the vertical extent of the applied averaging. The radial modulation in the accretion rate suggests that the inner disk edge recedes outward at a slow speed. The space-time plot further supports this notion, revealing that indeed the cavity is expanding outward ever so slightly. Initially, this trend is also seen in \citet{martel_lesur2022}, where nevertheless a stationary state is obtained at even longer integration times.

The primary objective of our present work is to explore cavity dynamics and investigate the transonic accretion velocities in that region. Given the added complexity from including the Hall effect, we argue that our 33 orbits of integration time at the transition radius provide a sufficiently long timescale for this analysis to be meaningful.

\section{Non ideal effects in the disk}\label{app:nonideal}

Here we showcase the non ideal MHD profiles for the cavity and the outer disk. We chose the \haa model and we plot the Ambipolar Elsasser number $\Lambda_{\rm A} = v_{rm A}^2/(\Omega \etaA)$, the magnetic Reynolds number $Rm = \Omega H^2/\etaO$ and Lundquist Hall number $\mathcal{L}_{\rm H} = v_A H/\etaH$ to demonstrate that indeed, our disk is dominated by ambipolar diffusion (see Fig. \ref{fig:r_diffusivities} below). At the same time, it is important to remember that the Hall effect is dispersive, not diffusive, and can be dynamically important even in areas that are AD dominated. This is shown in Fig 8, where we see that indeed in the cavity we expect the “strong Hall regime” to be present.

\begin{figure}[h]
	\centering
	\includegraphics[width=0.9\linewidth]{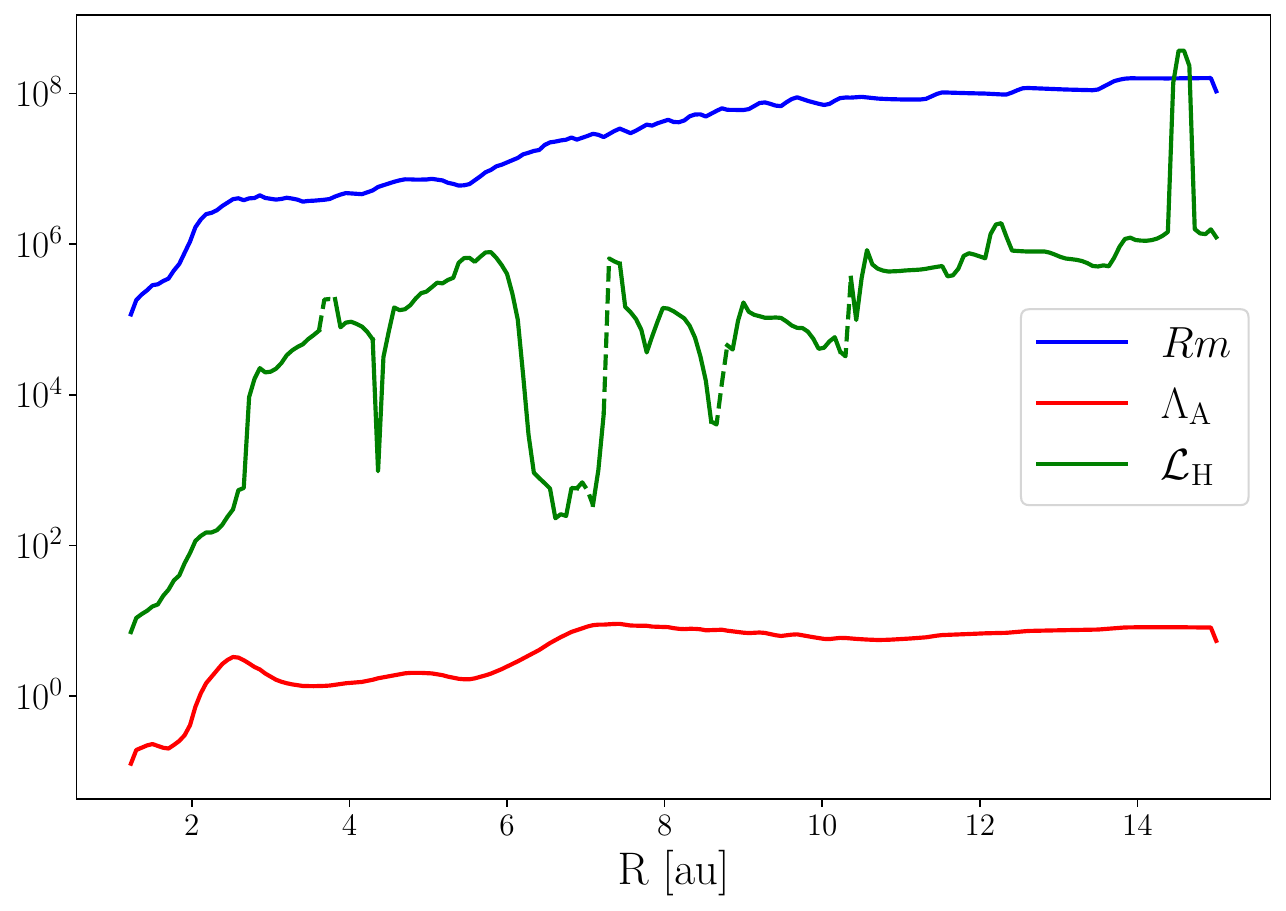}
	\caption{Time-averaged and vertically integrated profiles of the ambipolar Elsasser number, the magnetic Reynolds number, and Lundquist Hall number. We chose these specific dimensionless quantities as they are not dependent on the magnetic field strength and thus provide a better indicator of the relative importance of each of the non ideal effects – also see \cite{lesur2020}.}
	\label{fig:r_diffusivities}
\end{figure}

\end{appendix}

\end{document}